# Linking Administrative Data: An Evolutionary Schema

Jack Lothian[1], Anders Holmberg[2] and Allyson Seyb[3]

## 1 INTRODUCTION

The cross linking of disparate data sets across time, space and sources (administrative, surveys, censuses and registers) is probably the foremost current issue facing Central Statistical Agencies (CSA). If one reviews the current literature looking for the prevalent challenges facing CSAs, three issues stand out: 1) using administrative data effectively; 2) big data and what it means for CSAs; and 3) integrating disparate dataset (such as health, education and wealth) in order to provide measurable facts that can guide policy makers. CSAs are being forced to explore the same kind of challenges faced by Google, Facebook, Yahoo, which are using graphical/semantic web models(Ferrara, Nikolov, & Scharffe, 2011) for organizing, searching and analysing data[4]. Time and space (geography) are becoming more important dimensions (domains) as we start to explore cause and effect models. Central agency methodologists are being challenged to include these new perspectives into their standard theories, practises and policies. This paper presents a framework (or schema) for integrating and linking traditional and non-traditional datasets within a CSA. Like all survey methodologies, this schema addresses the fundamental issues of representativeness, estimation and total survey error measurement.

Over the past decade a new paradigm is being to emerge concerning strategies for integrating disparate datasets to provide new understandings of the data. The development of this paradigm is currently not focused and there are multiple paths of advancement being pursued: big data, evolutionary databases(Fowler & Sadalage, 2003), semantic web models, graphical query databases, and many others. All these areas of research have a core issue: linking large disparate datasets in a feasible and cost-effective manner. The complexity of the information, the fuzziness of the data inside each dataset, the fuzziness of the linkage strategies, the large number of disparate datasets covering disjoint populations, the lack of control of the content and quality of administrative data, and the size of the datasets precludes the use of many straight-forward classical solutions(Baker et al., 2013; Bakker & Daas, 2012; Holt, 2000; Zhang, 2012).

All the above-mentioned strategies appear to be following parallel paths to the same general solution. All these strategies are proposing viewing disparate dataset integration as an evolutionary process. The data and the database structure evolves as new information is added; as more datasets are added and linked; as new relationships between dataset are discovered and added; as new models of how different dataset interact are discovered; as new editing rules and methodologies are found; as the questions that we want answered change; as we become more knowledgeable of the data; etc. The evolutionary nature of the challenges implies that no fixed solution can succeed over an extended period of time. All of the above cited strategies embrace evolution and make it part of the solution.

The core design concept we are proposing is a simplified adaption of how Google and Facebook structure and search data. For Google, the databases are the files and their content on local internet servers. For a CSA the basic databases are structured collections of administrative files, censuses, and/or previously conducted statistical surveys. For Google, the point in time at which the measurements are made is not usually a relevant characteristic but in our strategy, it will be a fundamental aspect of the data.

---

[1] lothianjack@netscape.net
[2] Statistics Norway, anders.holmberg@ssb.no
[3] Statistics New Zealand, allyson.seyb@stats.govt.nz
[4] Graphical Queries Database (GQD) and Semantic Web technologies imply somewhat different paradigms. GQD defines a definition of index-free adjacency, meaning every element contains a direct pointer to its adjacent elements. The simplest GQD pointer is a first order tree of the one-to-one links. Second order trees contain all possible ways of linking 3 data sources. Semantic web links are first-order trees where the linkage function (the verb) becomes a generalized function.

Later in the document, we will also see that "space" or geography will become a necessary dimension of our design schema. Our schema will be underpinned and anchored by a space-time lattice through which our entities will travel. It is somewhat akin to the game called "Life". We will call the structured collection of common files (administrative, survey, registry or census) an evolutionary schema[5].

In the following sections, we will map out a conceptual schema for dealing with the integration of non-traditional and traditional survey datasets. It is important to note that we will be presenting a strategy for structuring, analysing problems and answering questions rather than a specific solution. As in classical survey design our final goal will be an estimation strategy. Current work in the data integration field tends to focus on just one phase of the process rather than the full process. The message that we wish to convey is that methodologists need to understand the complete process that will produce the estimates.

In the following sections of the paper, Sections 2 will present an overview of the database schema. Section 3 will discuss how estimation might take place in the evolutionary schema and highlight the importance of representativeness and registries. Section 4 will discuss measuring and controlling quality in this schema. Section 5 will discuss the importance of meta-data in the evolutionary schema. Section 6 will discuss the strengths and weaknesses of the schema and Section 7 will be a summary.

# 2 A STRUCTURED FRAMEWORK FOR USING "IT IS WHAT IT IS" DATASETS

## 2.1 "IT IS WHAT IT" IS DATASETS

In our evolutionary schema we will assume that all data sources integrated into the evolutionary database are provided by an outside agency that is beyond the control and influence of the owners of the evolutionary database. These outside sources could be administrative files, censuses, registries, client lists, commercial transactions, sensor reading, survey files, sample files, etc. Our source datasets will be what Sharon Lohr (Lohr, Hsu, & Montaquila, 2015) recently referred to as "it is what it is" datasets. "It is what it is" datasets are source files where the survey methodologist has no control over the selection probabilities nor the content of the files nor the quality of the information in the dataset. It should be noted that the true sample selection probabilities for the entities in these external sourced datasets may be non-probabilistic and/or unknowable. In addition, the quality of the information provided may be non-quantifiable or unknown.

As expressed by Sharon Lohr, the term "it is what it is" has a wider sense. As survey methodologists we may be asked to answer questions where the only source of information on this question is an "it is what it is" dataset. In this case, we may be forced to accept the limitations of the source and yet still try to provide a partial pseud-scientific answer to the question. In this case, it would be our sole source and it must be accepted as such. In addition, when using this source dataset, we need to be aware of the dataset's limitations or "what it is". Sharon talked about the need for the methodologists and the end users of the data to accept and understand the limitations of the source data. She stated that "it is what it is" datasets fundamentally changed our analysis paradigm and we needed to understand this point. In the following discussions the "is what it is" nature of the data sources will be an integral part of the schema.

---

[5] Here the term "evolutionary" refers to the fact that the database collects entities' event timelines and these timelines are always being updated with new current events. The event timelines evolve. In the paradigm of database design and programming, "evolutionary database" design has a different sense. In this paradigm it is the database design schema and algorithms that are always evolving in an incremental fashion. Our proposed design will be evolutionary in this sense as well.

## 2.2 THE "ELEPHANT IN THE ROOM" - REPRESENTATIVENESS

Unfortunately, there is a large elephant in the room when we deal with "it is what it is" datasets. The elephant in this case is the fact that they are not appropriate for making statistical inferences concerning the general population because the selection probabilities are non-probabilistic[6]. In a recent AAPOR task force report on non-probability sampling (Baker et al., 2013) stated that "approaches lacking a theoretical basis are not appropriate for making statistical inferences". It was further pointed out in an earlier AAPOR report (Baker et al., 2010) that statistical estimates and inferences drawn from "it is what it is" datasets cannot be trusted to be representative of the general population. In reference to the two AAPOR reports, Langer (Langer, 2013) quotes a well-known classical reference (Kruskal & Mosteller, 1979) stating that "We prefer to exclude non-probability sampling methods from the representative rubric." Langer is implying that one cannot ever claim that results derived from an "it is what it is" dataset is representative of the general population. This is a strong statement and raises questions about the ultimate usefulness of "it is what it is" data sources.

As the 2013 AAPOR reports states, the key is the risk associated with the source dataset not being representative of the general population. This is a serious risk because most "it is what it is" sources suffer from significant under and/or over coverage of various sub-populations within the general population. This is a systemic problem that is considerably worsened by cross-linking multiple "it is what it is" datasets. If one links a dataset that under-covers women with another that under-covers children, you get a linked dataset that under-covers both women and children. This is the Achilles heel of "it is what it is" data and if we cannot address this issue we will never be able to widely use "it is what it is" data.

As survey methodologists we must be able to defend ourselves from criticisms of bias caused by under-covering populations such as the under privileged or the rare populations. Without a methodology to measure under-coverage and correct its effects how do we maintain our credibility? Our schema offers a strategy for confronting this key issue.

## 2.3 CORRECTING NON-REPRESENTATIVENESS WITH REGISTRIES AND FRAMES

In our paper we address the key representativeness risk by:

- Assuming that we can create a stratification definition process that ensures that within each stratum we can assume that the observed entities were generated by a random process. Thus, within each stratum the observed entities are representative of the sub-population.
- We create 3 lighthouse registries and use them to measure under-coverage in various stratums. Then we use these registries to create calibrations (design-based designs) or models (model-based designs) or Bayesian priors (Bayesian designs) to correct for under-coverage.

The authors recognize that this strategy is naive and pseudo-scientific. It may not correct for all the biases created by the "is what it is" datasets over and under coverage. Yet it is a first step along a well-tread design-based path. The authors are following the historical path of development of corrections for census non-response. We consider this a first step in the correction for non-response bias. As we gather more expertise we will develop more mature and complex methodologies. While we recognize the risks of following a strategy that does not have a true theory behind it, we feel there is no other choice.

---

[6] One might think that this issue is only relevant for design-based methodologies, but it exists for Bayesians and modeller as well. In the model or Bayesian paradigms one must make assumptions about process that generates the records in the source datasets. Invariable these paradigms assume some sort of random generating mechanism from a super-population. Implicitly Bayesians and modellers are assuming a random selection process exists but is not controlled by the survey designer.

# 3 THE EVOLUTIONARY DATABASE SCHEMA – THE DATA MODEL

## 3.1 TIME AND CAUSE AND EFFECT RELATIONSHIPS

CSAs tend to view time as a descriptive characteristic rather than a fundamental dimension. Time becomes an estimation domain much like sex, age, race etc. Yet, whatever we survey, or measure evolves with a fixed time ordering and direction in time. Every observation is an event in time and when we combine two data sources we need to know how to order the combined events in time. Health, medical and most social scientists intuitively grasp this point because they are looking for cause and effect relationships or they wish to understand how the world is evolving. For these scientists, time is a transcendental variable that helps them make sense of any estimates. CSAs on the other hand tend to think of cross-sectional estimates (or panels) in time rather than a time series evolution. Time series analysis questions are usually "end of the line" analysis that do not affect the cross-sectional survey designs. Typically, during the survey design phase, time is considered irrelevant.

Time opens avenues for us to use, analyse and improve the quality of our datasets. Observing related events (a timeline of events) for an entity can provide us with a sense of the evolutionary changes in our data or the volatility of measurements over time. This can provide with proxies for measuring quality. Having a timeline of events for a common individual allows use to develop improved methods for detecting and fixing errors that are localized to one-time period. When one wishes to link entities and events in disparate datasets, the time lines can provide extra information that can improve the quality of the linkages and in some cases, it may allow us to develop quality measurement tools for the cross-linkages.

Time in our evolutionary schema is fundamental concept and every recorded event must have a time stamp. There is a time-ordering of all events in the schema, so we can distinguish between events such as diagnosis, treatments and results. Time can also open new avenues for improving editing, linking and quality measurements.

## 3.2 EVENT AND TIMELINE DATABASES

To illustrate how the evolutionary schema would work let us follow a simple example consisting of three administrative datasets: an annual filing of income tax returns; a collection of medical records from a group of hospitals, and a collection of school records from a group of school boards. We can view each of these three datasets as a list of unique entities (individuals in this case) and associate with each entity a set of date-stamped events. Each entity's record can be viewed as a timeline of observed events for that entity. Because new events will be constantly added to the database timelines, the timelines are always evolving in time. Hence it is an evolutionary database. The events are containers (holding virtual subsidiary databases) encompassing all the information gathered for this event. In practice, the event container might just hold a date stamp and pointers to secondary databases.

### 3.2.1 The annual filing of income tax returns with IRD

Let us call each collection of files that come from a common generating mechanism and frame unit a "timelines database". Thus, the collective of all the annual Inland Revenue Department's (IRD) filing of all New Zealand individuals through time would be the "timelines of individual annual IRD tax filers that submitted IR Form number 3 (or the IRD3_TL database for short). The data within this database would be structured in a specific manner. The fundamental unique key in the database might be the IRD number of an individual. Each annual IRD filing would be an event in the database and the event would be associated with the IRD number of a unique individual. Each event (annual filing) would have an event date that represents the measurement or collection date. Note there is no requirement that the data be collected on a fixed periodicity and entities' annual filings could be missing in some years. We will structure and conceptualize the IRD3_TL data as shown in the timeline in Figure 1.

*Figure 1: One Entity's timeline in the IRD3_TL database*

Entity's timeline

| Entity i | Event 1 | Event 2 | Event n |
|---|---|---|---|
| Entity ID | Event date | Event date | Event date |
| characteristic 1 | Event type | Event type | Event type |
| characteristic 2 | characteristic 1 | characteristic 1 | characteristic 1 |
|  | characteristic 2 | characteristic 2 | characteristic 2 |
| measurement m | measurement k | measurement k | measurement k |

Each timelines database will be of a collection of the timelines for the entities covered by the IRD3 data source files. Thus Figure 1 represents a timeline while Figure 2 represents a timelines database.

*Figure 2: The IRD3 Timelines Database for time period $t$ to $t + T$*

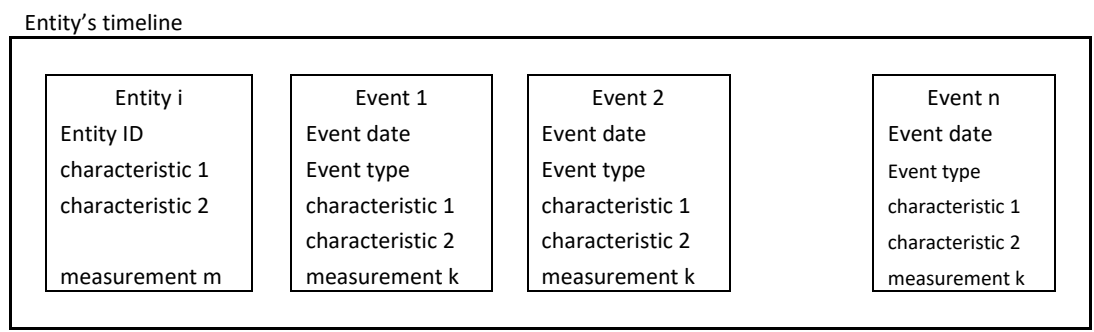

Within each database the collection of entities must be of a common type, but each timelines database could refer to the timelines of a number different entities and events: persons, families, businesses, institutions, dwellings, pets, etc. Using survey terminology, there must be a common sampling or frame unit within each timelines database. An entity's timeline record is made up of a common entity ID (key) plus a collection of events that were observed for that entity at specified event dates. Each timeline grouping will contain at least one event, but multiple types of events may be observed. Again, the timeline of events could be containers with pointers to subsidiary databases.

Maintaining the IRD3_TL database would be straight forward, minimalistic, cost effective and fast. Whenever a new batch of annual filing come in, one only needed to find the associated IRD_ID in the database and add a new event to the record. The evolutionary database could consist of containers of pointers to other subsidiary databases rather than data values. In this case one would only need to update the pointers. If the database was structured properly, this activity would require minimum re-indexing and sorting. Once a new batch of records was appended to the end of the subsidiary database it might never be touched again.

### 3.2.2 The structure of the IRD3 timeline database

Figure 3 shows how one might create an evolutionary database within a previously established database environment containing standalone yearly IRD3 databases.

*Figure 3: Constructing the timelines database from subsidiary databases*

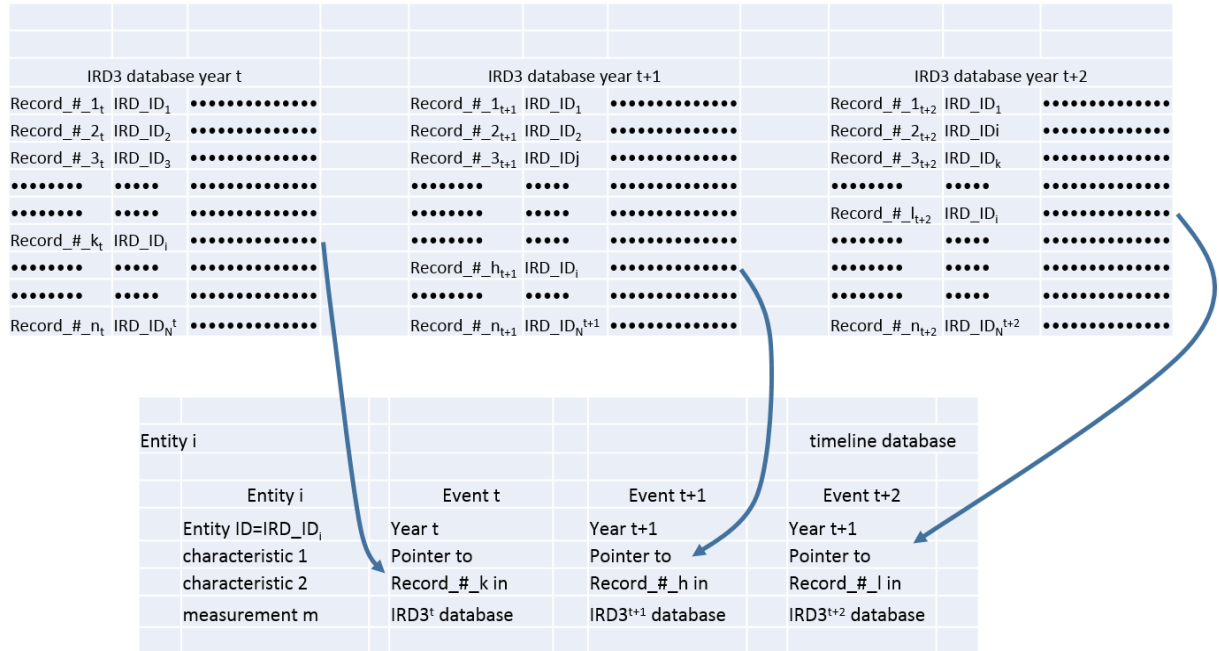

The IRD3 databases are assumed to be previously created databases that where developed under an alternative survey design strategy. The imposition of the timelines schema onto the IRD3 databases does not require the owners and previous users of the IRD3 databases to change any of their methods. As long as the standalone IRD3 databases remain static, no changes will occur in the timelines database. Even edits of the standalone IRD3 datasets may not require any updates to the timelines database. Additions, deletions or changes involving an IRD3_ID will definitely require a recompilation of the groupings inside the event receptacles. Adding new years of IRD3 data will add events to the timeline database and require an extra compilation of the groupings for that database.

### 3.2.3 The collection of health events from a hospital

In this case, the unique key in this database might be the internal patient number used by the hospital to designate patients and the events could be diagnosis, treatments, tests, etc. Unfortunately, inconsistencies might occur causing some patients to be assigned multiple unique keys. This may be problematic if the numbering system for the keys changes over time or between health providers. In this case the timelines of the entries might be fragmented in time. The long run solution for this is to make the unique key an internal database key that can be remapped so as to join up the fragments. The strategy for joining broken fragments could be a future evolutionary development.

In this timelines database (Health_TL) multiple different types of events might be recorded; each being pulled from a subsidiary database. Again, the records must all relate to a common entity; a patient and there must be a mechanism for grouping a patient's event timeline. Inconsistencies in this type of database could be significant and over time it may be necessary to incrementally put in place mechanisms to eliminate the inconsistencies. Thus, database design and programming would be evolutionary.

### 3.2.4 The collection of education events from a school board

For the education timelines database, the identifier might be a school board student assigned number while the events might be marks on specific tests, special educational indicators, discipline reports, etc.

Each event would occur at a specific time and there would be attributes defined for each event. Some attributes would be defined at the identifier level (like birth date, last address, name) while others would only be defined at the event level (like date of transaction, observed attribute, etc.). Each event might observe different attributes.

In this timelines database (Educ_TL) multiple different types of events might be recorded; each being pulled from a subsidiary database. Again, the records must all relate to a common entity; a student and there must be a mechanism for grouping a student's events into a timeline.

Inconsistencies could be a serious issue with educational data, especially if it is being grouped with disparate events widely separated in time. Changing schools could lead to the issuance of an alternate student number and this could cause fragmentation of the event timelines. Children tend to leave home and change their addresses when they graduate, and students' names can change significantly between childhood and adulthood. The database design and programming would be evolutionary so that mechanisms can be developed at future dates to resolve these inconsistencies.

## 3.3 LINKING OR RELATING TIMELINE DATABASES

Users of this evolutionary schema will want to define and discover relationships (cross-linkages or connections of disparate timelines) between the timeline databases. They may want to explore the relationship between poor school performance and health or between health and wealth. To achieve this objective, one must start at one of the timeline databases and connect to the other timeline database through a linkage relation database for these two datasets.

*Figure 4: Functional relationships in the evolutionary system*

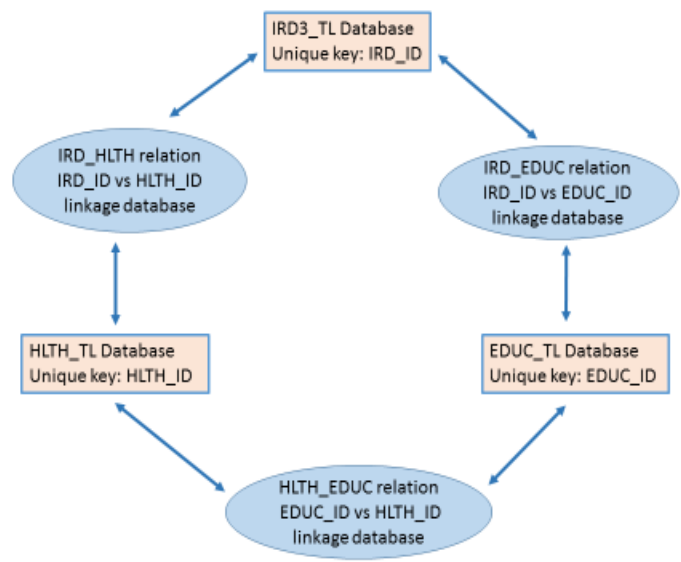

Figure 4 illustrates how the potential relationships (linkages) between the timeline databases are defined. All linkages are one-to-one and are not necessarily exhaustive. Direct relationships (linkages) only exist between two timeline databases[7]. Linkages involving three or more timeline databases only exist virtually or indirectly and the solutions are path dependent. Thus, if A, B and C are three timeline databases then $(A \cap B) \cap C \neq (A \cap C) \cap B$. The intersection of the 3 databases is path dependent neither commutative nor transitive. Anyone who has observed Google searches is aware that the results of the search depend on the order of the words used to do the search. This will also be the case in our evolutionary database.

---

[7] We are assuming that first-order nearest-neighbour effects are dominate.

Like Google, how we define complex higher dimensional relationships between more than 2 timeline databases is by "stepping-through" simple one-to-one linkages that only inter-relate two timeline databases. These kinds of linkages (or searches) are sometimes call sematic web or graphical databases searches. In these types of database, linkages are always local in the sense that they only inter-connect two timeline databases and longer paths are defined by moving through the local linkages.

Our database design is evolutionary in the information technology (IT) sense. The algorithms linking, editing or transforming the data will incremental evolve as more knowledge is acquired. Initially some linkage relationships might be undefined and implemented on a need to have basis. Fields in subsidiary databases, events, timelines and new survey data sources can be iteratively added as they become available or are needed. The maintenance of the evolutionary database is devolved and distributed amongst local groups with strengths and experience in the local data. The timeline databases can be disseminated amongst unrelated control groups. The group of individuals responsible for the standalone Health datasets need not have any contact with the IRD3 team or the education datasets team. In addition, a separate team could be assigned responsibility for creating and maintaining the linkage databases. Each team could add events, edit fields independent of the other groups. New linkage technologies could be implemented without requiring any revisions to the timeline databases or their subsidiary databases. It becomes a distributed and cooperative evolutionary system where local changes will not force a recompile of the complete system. An added benefit is that the evolutionary network of databases and linkage functions will have natural buffers that will tend to isolate the effect of local failures.

## 3.4 TIMELINES ARE A FUNDAMENTAL CONCEPT

In the data model presented, time is a foundational concept. Our intent from the beginning was to design a database that could relate causes and effects that were dispersed over time, so we can answer such questions as: Does education early in life effect wealth later in life? Does education early in life effect health later in life. Etc. We anticipate that time will be a fundamental concept in estimating cause and effect relationships. A preliminary and necessary requirement for doing cause and effect analysis is ordering events longitudinally into a timeline.

The linking of events into a timeline can open avenues for improved linkage strategies. Missing linkage variables can be estimated from other events in the timeline and inconsistencies in names, addresses, age, etc. can be edited and standardized by analysing the full timeline. The linkage strategy could depend on time vectors instead of a single value. Perhaps this kind of linkage strategy could help us deal more effectively with name and address changes observed over time.

As we will see in the next section, space (geography) will also be a foundational concept. Both space and time will become rigidly anchored in our schema for all entities in our evolutionary database. Entities (firms, patients, persons, students, institutions, etc.) can and will move around on this two-dimensional fixed grid and events will occur on one of these grid points. Of course, this is a theoretical paradigm where we always assume that space and time exist and are known absolutely in some sense. In practice uncertainty can exist in the measurement of space and time and either variable may on occasion be missing. Yet it would be helpful in discussing our schema if we initially accepted that time and space are always known facts.

## 3.5 THE GROUPING OF EVENTS WITHIN A TIMELINE

In our schema, the grouping of events based upon a common entity ID is a deterministic function and not a linking process which we see as a fundamentally probabilistic function. In our terminology. The creation of the timeline in Figure 1 will be referred to as a "**grouping**" function while the blue ovals in Figure 3 will be referred to as a "**linking**" function. Entity IDs are assumed to be known true facts.

Empirically, we may have time fragmentation caused by issuance of multiple IDs to the same entity or inconsistencies caused by the issuance of the same ID to multiple entities but for the purposes of this paper we will assume these inconsistencies do not occur. In most cases, this should be a reasonable approximation of reality, but perverse data sources may exist. Resolving these types of inconsistencies could be a future evolutionary step.

Linkage functions are assumed to be probabilistic in the sense that the linkage function always has significant uncertainty associated with each identified link. Linkage functions (the blue ovals) always produce a sub-sample of the two source datasets with an underlying probabilistic element. The number of linkages found and the "truth" of each link is probabilistic (random) in some sense. Thus, we see each output linked dataset as a probabilistic sample. While it is almost certainly true that our output dataset is a probabilistic sample, we have little knowledge concerning the selection probabilities. We are not even certain if the selection is with or without replacement. A key issue is whether the probabilistic sample is representative of our target population. What we will assume is that sample selected will be non-confounded (at random) within some estimation domain.

## 3.6 CONVENIENCE SAMPLING

Our linked or integrated output dataset will be an opportunity sample[8]. Our integrated sample is a sub-population derived using relationships and networks to which we have access and we have no control or knowledge of how these relationships were constructed. Researcher using this sample cannot scientifically make generalizations about the total population from this sample because it may not be representative of the target population. Strictly speaking, convenience samples are non-probability samples, yet we can hypothesis a hidden underlying probabilistic selection mechanism that is random within some estimation domain (stratum). Thus the non-probabilistic element of the selection process only effects the balance between estimation domains. The credibility of a researcher's results when using this hypothesis will depend on convincing the reader that the researcher has properly compensated for the imbalance between domains and the final estimates are representative of the population of interest.

# 4 ESTIMATION WITHIN AN EVOLUTIONARY DATABASE

## 4.1 ESTIMATION PLAYS A CENTRAL ROLE IN THE DESIGN OF THE EVOLUTIONARY SCHEMA

Up to this point, our evolutionary schema concerned itself exclusively with data models and database designs that exploited the scalability, flexibility and efficiencies of an evolutionary data model. Yet, the objective of any survey is estimating some collection of parameters or effects or relationships and the evolutionary schema must ultimately support this function. Our objective is to use estimates derived from this schema to understand real world populations and how entities in these populations interact in time and space.

In order to interpret estimates from the data we have, we must obey some principles and stick to some theory. From survey theory we can borrow three different principles which we can apply. The design based or randomization theory (e.g. see Särndal, Swensson and Wretman, 1992) is one principle which emphasizes that attribute values of the records in data are fixed values and that it is the random selection of the elements in the dataset that provides the representativeness of the target population through a sampling frame and this also gives us the means to assess the accuracy of estimates. Another principle is the predictive or model based approach (see eg. Valiant, Dorfman and Royall, 2000), then the values are regarded as realizations of random variables and the design by which the survey elements are selected is of less importance. A third principle that can be implemented is a Bayesian inference framework. It has been put forward as useful for analysis of small non-probability samples and appear to be a direct alternative when data from different sources are being combined (see e.g. Rao (2011) and Little (2012)).

---

[8] Note that we are not referring to our integrated sample as an "it-is-what-it-is" dataset. That term we will reserve for source datasets that are not under the control of the methodologists responsible for doing the estimates.

The choice between these principles are in spirit based on inferential philosophy preferences but tends to depend on the willingness to accept and depend on statistical models, the trade-off between the statistical properties bias and variance and the degree of how context specific the objectives of the statistics are. If on one hand the goals are descriptive inferences for a large number of finite population characteristics and avoiding publishing biased estimates, the design based principle has been a traditional choice, if on the other hand we are seeking analytic inferences that are focused on causal mechanisms, then the model dependent principles appear to be the natural paradigms.

In the literature to date, there has been more discussion of the potential data models rather than how estimation enters the schema. We believe that many of the discussions to date have missed the central role that registries and frames[9] must play in estimation. Registries and frames are the support scaffolding for estimation done within the evolutionary schema. Without this scaffolding our estimation strategies will be weak and prone to failure irrespective of the principle we use.

## 4.2 REGISTERS/FRAMES ANCHOR THE EVOLUTIONARY DATABASES

Our evolutionary schema may include events and timelines databases from subsets of the targeted population of entities with unknowable selection probabilities. In addition, the reporting of events may be inconsistent between timeline entities. Using survey methods terminology, we have no idea of how much of a given target population our data covers.

In the three examples of timelines database given, all three databases include entities from what we may regard as one super-population: the union of all residents of NZ during a fixed time interval defined by the union of all the timeline event dates. This super-population is a virtual population because it may contain entities whose timelines never overlap because they lived in disjoint time periods (for example, one died before the other was born). In general, the union of all the entities in an evolutionary database will be an abstract set of entities that does not conform to any real-world population. In survey terminology, our evolutionary database may cover a mixture of entities from disjoint and unrelated populations and registries created from timeline databases may contain out-of-scope entities and miss significant numbers of our in-scope (or target) population[10].

In the three database examples given, our evolutionary database covers students, patients, and tax filers from multiple time periods. While all three entities are "people" who resided in NZ during some time period they are probably a collage of real world populations. For example, some might be different categories of foreign residents or visitors, some might have died or left the country before others were born or arrived. From an estimation point of view, we may not be able to know if a selected entity was in the target population we are studying.

It is as if our database included a random sample of "apples & oranges" from two populations of fruits with unknown population sizes. To further complicate the situation each sampled fruit is wrapped in paper so we did not know what population it belonged to. Our stated objective was to do various estimates of characteristics and quantities of these apples and oranges in the original populations. In this problem we do not know the sizes of the two target populations and we do not know which sampled unit belongs to which population. Under such conditions, estimation for the two distinct real world populations is not possible. If we wish to do estimates, some essential information is missing from our database.

---

[9] In this paper registries are any list of unique entities involved in any of the various timeline databases, while frames are registries that are an empirical realization of a specific target population.
[10] These two concepts are sometimes referred to as over-coverage and under-coverage.

The extra information that we need to do estimation is an entity frame $U_F$ for the population of interest $U$ (i.e. the target population). Depending on the information available, an entity frame can be derived from a statistical (core) entity registry. The statistical entity registry may be an integrated, micro-merged and maintained list of entities created from the combination of different administrative data sources based on identifiers that are unique to the various data sources. In addition, we need a linkage relationship database that cross-links the entity registry and the entity timeline dataset of interest. With this additional information, we could calibrate the selected linked sample to the frame[11]. This can be done to achieve coherence between known information in the frame and the linked dataset. By using aggregate data from the frame such as domain totals $X_d$ and domain counts $N_d$ and regard them as known, weights and calibration equations can be constructed that reproduce these totals from the linked dataset. Hence, we construct weights $w_k$ that satisfy the principal expression:

$$\sum_A w_k \mathbf{x}_k = \sum_{U_F} \mathbf{x}_k = \sum_U \mathbf{x}_k \qquad (1)$$

where $A \subseteq U_F$ is a subset of elements in the linked dataset, and $\mathbf{x}_k$ is a vector of known variables given from the frame. Depending on the circumstances different versions of equation (1) can be used. For simplicity we assume that $U_F$ is an extraction from an entity register that perfectly represent $U$. To cover domains, we just add the subscript *d* and define variables

$$x_{dk} = \begin{cases} x_k & \text{for} \quad k \in U_d \\ 0 & \text{for} \quad k \in U - U_d \end{cases}$$

where an important case of $x_k$ are indicator $1_k$ variables that gives us, $N_d = \sum_{U_F} 1_{dk}$.

The estimation of other parameters e.g. an unknown total $Y_d = \sum_U y_{dk}$ from the linked dataset is then done by using the same set of calibration weights, i.e. $\hat{Y}_d = \sum_A w_k y_{dk}$.

This is a popular technique often used in a design based inference often using the same weights for all estimates (see Särndal and Lundström (2005), Särndal (2007)). To apply it here we have to make the naïve assumption that the linked sample within an estimation domain is a non-confounded[12] sample from the target population. Then (design-consistent) calibrated estimates are possible, and under this simple scenario one could estimate variances.

If we have entity level information from the frame, we can take the calibration technique one step further and apply explicit models, i.e. compute calibration weights that instead of $Y_d$ and $N_d$ a reproduce model predicitons of superpopulation parameters $Y_d^*$ and $N_d^*$ (see Wu and Sitter (2001)). In this case a separate model can be applied for every study variable and domain although that would need considerable modelling effort.

In a Bayesian framework we can denote the information we have from the entity frame by $Z$, it would enter in the specification of the prior distribution of the population values, $p(Y|Z)$. And it would also be used as covariates during the generation of the posterior distribution and parameter estimation when the prior is confronted with the linked data.

---

[11] In Bayesian terminology these calibrations would be prior constrains on probability distributions.
[12] Non-confounded might be considered as synonymous with "observations are missing at random". Thus we assume that the observed sub-population is representative of the full target population in every variable. Non-confounded has a wider contextual meaning that implies the measurement variable is an unbiased estimator in both the statistical and non-statistical sense. The term "unconfouded" is discussed by Lee. H., Rancourt. E. and Sarndal, C-E. (2001) Variance estimation from survey data under single imputation. In R.M. Groves, D.A> Dillman, J.L. Eltinge and J.A.R. Little (eds) Survey Nonresponse, New York Wiley.

## 4.3 LINKING TIMELINE ENTITIES TO THE FRAME

Ancillary information is necessary for making unbiased estimations from the convenience sample obtained from linking two datasets. To illustrate this point, we will demonstrate how estimation might occur in a simple example. We will be estimating the relationships between heath and wealth of individual entities in the current New Zealand population (the top-right blue ellipsoid in Figure 3). There are multiple difficulties with these datasets. First, the IRD3 and HLTH timeline databases contain out-of-scope units and have significant under coverage of the target population. In survey terminology terms, we are neither sure of the true target population size (big $N$) nor the true sample size (little $n$). Without some knowledge of the true $n$ and $N$, how can we make unbiased estimates of the relationships? How do we answer such questions as; how many children are expected to have a specific health issue? At best we can estimate ratio or proportional effects that apply to unknown target populations. Second, we know that both timeline databases are confounded, possibly in different ways. Thus our timeline databases are not representative of the target population. In general, the poor, immigrants, the very young, the very old, people with handicaps, stay-at-home mothers, etc. may be missing from one or both of the timeline databases. In survey language terminology, our linked sub-sample is a biased sample. However, if assume that the sample within each estimation domain *d* is "at random" or non-confounded and if the entity frame is of good quality giving us $n_d$ and $N_d$ we can apply the calibration technique and improve the estimates by decreasing the bias of estimates. Without the frame we would not have this possibility.

### 4.3.1 Linking the integrated sample to frame

There are two strategies one might use to link the integrated sample to the empirical frame. If we are fortunate enough to have quality information available to link the sampled records at the entity level, we can create a micro-entity estimation file with weights applied at the entity level. Even if entity level linkage is not possible we could link at the domain estimate levels if each timeline database file contains the domain stratum identification variable. We will go through both cases. In each of these simple cases being considered we will be making the naïve assumption that within an estimation domain the data is non-confounded.

### 4.3.2 Estimation when entity level linkage with the frame is exists

To illustrate this point let us look at the simple case of providing estimates for a relation between two of our timeline databases: heath and income. In addition, let us assume that we have a registry of current residents of New Zealand (perhaps derived from a recent census). Furthermore, a linkage relationship exists connecting the resident ID (or census ID) on the frame to the person's personal IRD number (IRD3_ID) in the timeline database. Our evolutionary database only contains first-order links, so we must step through the two linkage relationships. Figure 5 shows one of the possible solution paths.

*Figure 5: Estimation processing steps when entity level linkages exist*

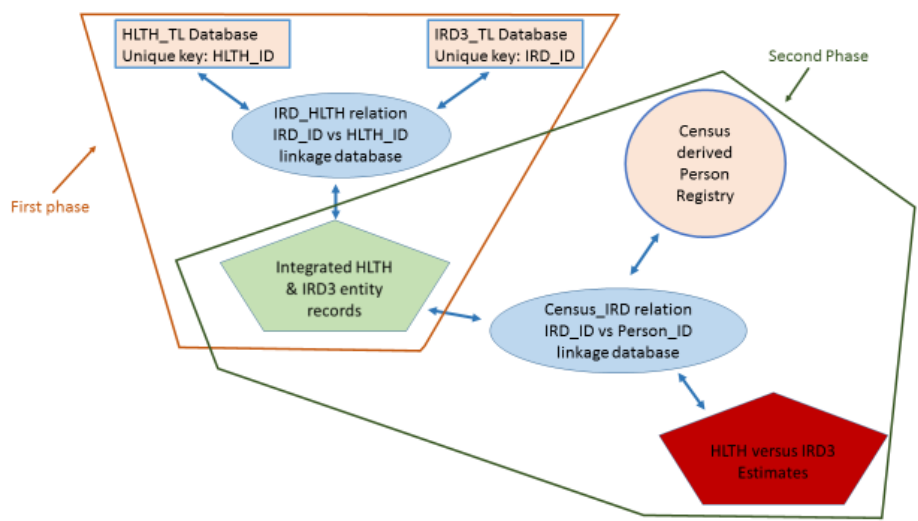

Each blue ellipsoid in Figure 5 could be considered indicative of a survey processing step or a linkage-step and the steps through the linkages we will call phases. Figure 5 illustrates what a phase means in our schema the green pentagon is our output integrated dataset from phase 1 and it comprises two container fields holding the IRD3_ID and the HLTH_ID. Without context the output sample is not generally sufficient information to derive reasonable estimates. In phase 2 the context (or calibration) is provided that turns the phase 1 output file into estimates. In the pursuit of acquiring and increasing the use of administrative data the current discussion in CSA often lack this second contextual phase. We feel this is a key point, and that frames and registers can provide the scaffolding that supports quality estimates. In this particular example, hopefully a statistical entity register of persons with small coverage errors can assist in improving estimates based on the linked data.

### 4.3.3 Estimation when entity level linkage with the frame does not exists

In some cases, no reliable entity level linkage information may be available, or it is possible that we wish to link disparate information collected from different but similar entities. If the three datasets (health, wealth, census frame) are non-confounded within an estimation domain, concurrent and have common domain stratification variables, then a design-consistent estimator may be found, and calibration weights can be calculated that correct for biased inclusion rates between stratums. Figure 6 illustrates this case.

*Figure 6: Estimation processing steps when only common stratum information is available*

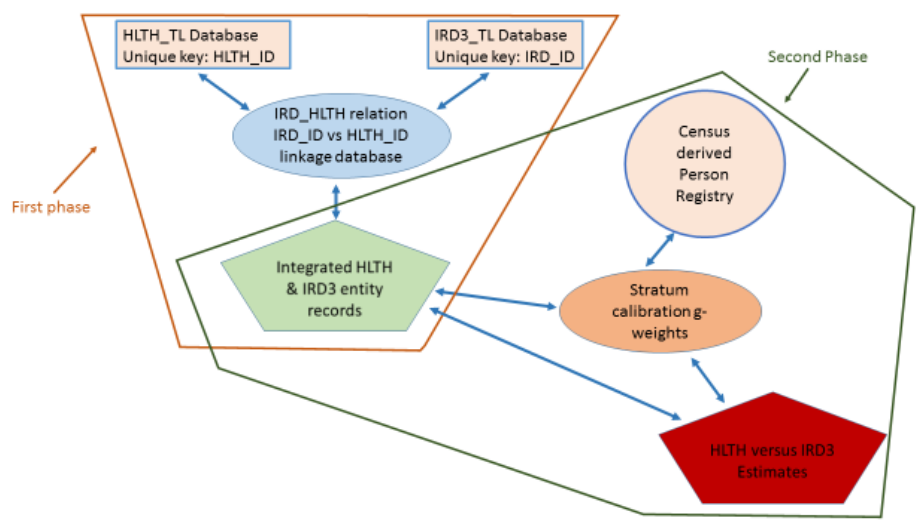

In Figure 6, the linkage between the integrated entity level data (green pentagon) and the frame (beige circle) will be indirect. An intermediary file will be created (the orange oval) by combining information from the integrated sample and the frame. Design-consistent weights for the estimates will be the inverse of the sampling function $\pi = N_d/n_d$, where $N_d$ is the target population size (big N) derived from the frame $n_d$ (small n) is the sample size derived from the integrated file. Estimates (red pentagon) will be calculated by applying these weights to the entity level records in the integrated file. This is a simple and naïve example showing how we might a use a frame to improve the quality of our estimates when entity-level links do not exist.

Figure 6 illustrate the case when the linkage between the entity level data (green pentagon) and the frame (the circle) will be indirect. The orange oval represents the combined information from the two. As an example, assume the linked timeline databases contain categorical data such as a geographical (domain $d$ with $D$ categories) and a socio-economic classification (domain $d'$ with $D'$ categories) and that we have this information in the frame as well but not necessarily simultaneously. With $\varsigma$ being the domain category indicator we can form $\mathbf{x}_k = \left(\varsigma_{1k}, \ldots, \varsigma_{dk}, \ldots, \varsigma_{Dk}, \varsigma_{1'k}, \ldots, \varsigma_{d'k}, \ldots, \varsigma_{D'k}\right)^T$ and the right hand side of (1) will be the domain counts $N_{d\bullet}$ for $d = 1, \ldots, D$ and $N_{\bullet d'}$ for $d' = 1, \ldots, D'$. Hence, we use the marginal distribution of the domain categories from the frame to support the estimation. The weights that satisfy a calibration equation will depend on the inferential principle used and with this $\mathbf{x}_k$ there is no nice expression, however numerical computations are not difficult.

Figure 6 presents the simplest case, but we could add complexity by adding more source data files, linkage functions or we may wish to relate information on the frame to the integrated file. In this example. The estimation function (orange oval) is design-consistent estimator but we could replace it with a Bayesian estimator (the orange oval would be the priors) or model-based estimator. Figure 6 should be interpreted as a generic strategy. The key point is that ancillary information is required to remove coverage and bias problems associated with the input sources.

## 4.4 REPRESENTATIVENESS OF THE LINKED SAMPLE

Registries and frames are the scaffolding that supports quality estimates. In order to make meaningful estimates our linked sample must be representative of the target population. Intuitively, most humans have a sense of the Law of Large Numbers. We know that small observation sets are untrustworthy. So it natural to assume that adding more data to our system has to make it more trustworthy but that is not true in a linking situation. The Law of Large Numbers is an additive concept, where we are averaging. On the other hand, for linked datasets it is a multiplicative situation. To demonstrate this problem let us return to the example we used in Figure 5, where we are linking health and wealth data sources, so we can explore how education affects wealth. Let us make a few simpler assumptions. We will assume that each dataset suffers from under-coverage. Perhaps, the IRD3 data does not cover some the stay-at-home-wives and the health data under-covers children. Figure 5 illustrates what happens when you link the data. The integrated dataset[13] (green oval) acquires the weaknesses of both source datasets. Integrating further dataset sources only makes things worse. The output dataset will acquire the weaknesses of every input source. When dealing with integrated datasets one should always assume the output linked dataset will not be representative of the target population. Estimation strategies must be developed with this fact in mind.

*Figure 7: Coverage biases in integrated datasets*

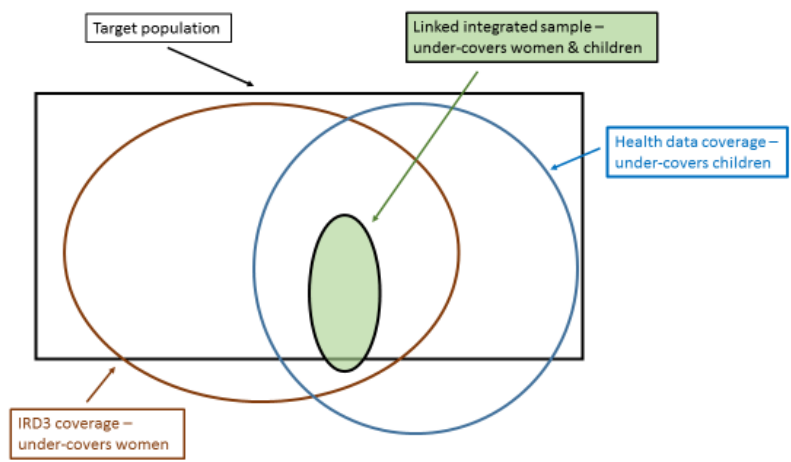

Many discussions on using linked or administrative data ignore two critical questions. How are we going to do estimates and will the derived estimates be representative of the group of entities that interests us? These are tough questions but disregarding them will ensure failure. Too often, practitioners seem to assume that adding more and more disparate source datasets will somehow solve the problem, yet Figure 7 shows that this is not generally true. In a sense this may be partially true, but the required ancillary information must suit a specific purpose; it must allow one to construct a frame that enumerates the target population by the required stratification variables. Thus, a central design characteristic of our evolutionary database is the existence of registers or frames that help researchers to make estimates that are representative of the true population. We must know the weaknesses of our data sources and have estimation methods available that minimize the effect of these weaknesses.

---

[13] In this section, the term integrated dataset is synonymous with the term linked sample. The integrated dataset is a sub-set of the intersection of the two data sources. It is the entities in the intersection that were also linked.

## 4.5 REPRESENTATIVENESS CAN BE MORE ESSENTIAL THAN EXACT LINKS

The current literature give the impression that finding exact matches between entities in disparate datasets should be the overwhelming imperative when integrating disparate datasets. Yet, without a universal entity ID that crosses disparate datasets, the so called exact matches are often matches of entities with similar names and addresses and not necessarily exactly the same. Exact matching without a universal ID is often of poor quality without field standardization, introduction of aliases and blocking. In addition, matching often introduces fuzzy concepts such as soundex matching that introduce uncertainty into the matching process. Calling these matches exact can give a misleading impression of the data quality.

Our perspective is that maintaining representativeness should take priority over obtaining exact matches. Let us consider two cases. In the first case, we match a family in one data set with a family in the second dataset. But we constrain the match so that the two entities must live in the same neighbourhood with a similar priced house and they must have families that demographically match (for example the same number of children of the same ages, etc.). After matching we discover we matched the Smith Family in one dataset to the Hodges family in the second. In the second case, we constrain the match, so the two entities must have similar names and addresses. After matching we find that we have matched the Jim Smith family to the J. Smyth family. In addition, we note that the Smith family has two children while the Smyth family appears to have no recorded children.

In this case, we believe that better estimates would be obtained by matching families with similar demographics and wealth than by matching on similar names and addresses. We are not arguing against name matching but rather suggesting that the concept of linked or integrated be generalized and representativeness of the integrated sample should be the primary concern of the linking process. In some sense, we making a Bayesian type assumption.

## 4.6 CREATION OF REGISTRY FILES CAN BE EVOLUTIONARY

Sometimes, creating a representative sample file may require using multiple registries. As an example, let us consider our example in Figure 6. While the integrated (linked) sample file collects all the weaknesses of its sources, perhaps together they might give us the calibration information (or priors) that we need to eliminate representational effects in our estimates derived from these biased datasets.

If we can create a unique union of the IRD3_ID and HLTH_ID entities, then we can create a registry that quantifies the under-coverage of women in the integrated dataset. We can go one step further and create a union of the entities in all three of our example sources. The union of IRD3_D, HLTH_ID and EDUC_ID would then quantify the under-coverage of both women and children in the integrated dataset. Creating this grand-union of entities is a challenging task fraught with many problems but we could tackle the frame creation in an evolutionary manner, where we start off with a naïve algorithm and as we develop more expertise we would develop more sophisticated algorithms. The key issue is defining our target population and identifying an estimation strategy that adjusts for the known biases in the source data.

## 4.7 THERE ARE FOUR KEY DIMENSION: TIME, SPACE, PERSONS, AND FIRMS

The structural supports for our scaffolding will be the three traditional entity frames used by CSAs. These are dwellings (location), persons and firms. Traditionally, the census played the role of the dwelling and person registries while the business register provided a firm registry. To these three key registries we add time so that cause and effect relationships can be estimated. Like in science, time is a special dimension with unique properties quite different from our other 3 entity dimensions. Yet never-the-less it will enter into many estimation problems.

How we see this operating is similar to major support columns and beams holding up the estimation structure. These three frames will be the starting point for adjusting for non-representativeness effects. The three frames underpin the creation of all the intermediary registries/frames that we might develop. In our example of health versus wealth, perhaps we postulate an effect where wealth gives patients access to better hospitals. So we are looking to see if evidence of this effect exists. Here we might use the business register to give us a frame of all hospitals and we might use the census to estimate the patient frame. In the process of estimating the existence of this effect we may create subsidiary frames only covering our targeted hospitals and patients.

# 5 A TEMPLATE FOR CREATING A CORE REGISTRY FROM ADMINISTRATIVE DATA

## 5.1 ESTIMATION AND REPRESENTATIVENESS REQUIREMENTS IMPLY THE NEED FOR REGISTRIES

The evolutionary schema's estimation strategy is buttressed by our three core registries: firms, persons and land. Ultimately, estimation must be the objective of any central statistical agency output. We cannot perfectly enumerate entities nor the attributes that we associate with them. In addition, many of the relationships that we are trying to illuminate and understand are complex latent effects that are not directly observable. All data released by CSAs are the result of some estimation process. For the authors estimation is the primary goal in our schema; all other objectives are sub-goals necessary to do estimation effectively.

Estimations must be grounded within some bounded dimensional space. In general, estimations are only valid for some local space or population. With administrative data sets the observed entities (members of our population) are often not representative of the full population of interest. Some datasets do not include children while others only include children. Similarly, other datasets may under or over represent the rich, poor, immigrants, native-born, urban, rural, male, female, old, young, sick, healthy, educated, uneducated persons. Most source administrative datasets are not representative of the full population of interest. In statistical terminology our observed response set is biased.

When we create a union dataset of all the entities covered by multiple administrative dataset sources, the resulting population size is often an order of magnitude larger than the current estimated population of entities. Alternatively, when we create an intersection dataset for the available administrative data sources, the coverage of the population rapidly plummets as more administrative datasets are joined. Typically, the entity count in an intersection dataset is an order of magnitude less than the current estimated population of entities. Thus the authors believe that must recognize representativeness as the core issue in our schema. The data sources "are-what-they-are" and we accept that fact and deal with it the best we can. Without some strategy for dealing with this issue any estimation process will be of poor quality.

Our answer is to create core registries that that will allow design and model-based methodologists to calibrate the estimates to correct for non-representativeness and Bayesians can use the core registries to get more effective priors. In general, we hope that if we correct for key demographic, geography and economic effects the results will be representative (unbiased) of the population of interest. At the very least the core registry will allow one to correct for the standard non-response effects. We are treating non-representativeness corrections as being equivalent to non-response corrections in classical survey design.

In the next section we will outline a possible strategy for constructing a core registry. There are commonalities of function and design that cross the three core registers defined in our schema. Each registry will be made up of well-defined entities that exist in the real world and are theoretically finite in number and countable. In each case the registries will cover a wider population of entities than the current active population. In addition, there will be multiple sources of varying quality indicating whether an entity existed or not and over which period of time. Entities will have birth and death dates. Hierarchical structures may exist within each registry (firms: enterprises, establishments, locations; persons: households, families, persons; and geography: regions, land holdings, buildings, addresses).

## 5.2 THE TEMPLATE

The most mature and best understood registry is the business register (BR). Most central statistical agencies maintain a business register that is remarkably consistent across agencies. The next most mature registry is the personal registry but typically its design and maintenance is simpler than the BR, often being equivalent to the census statistical sampling frame. The least well-articulated registry is the land registry. An actual land registry may not exist and when it does it might be either a list of contact addresses or grid cells drawn on a map.

The basic BR template is mature, well understood and supported by a broad international consensus. Thus we will use a simplified BR structure as a template for illustrating how one might construct and maintain the three core registries. In the following discussion we will use this BR template to construct a working model for the less well-defined person registry.

The business register is not a core registry in the sense that we define it. The current standard design of the business register encompasses a complex system of interacting files rather than one core list. In our paradigm there are three basic types of files encompassed within the business registry system: the source **administrative registries**; the full **entity registry** and the **statistical registries**. In the following discussion we will explain this terminology.

The Stats NZ business register system enumerates all NZ business entities that have shown any indication of economic activity in the past $\Omega$ years. Similarly, our person registry will enumerate all persons who have shown any evidence of NZ residency during the past $\Omega$ years. Both registries cover non-corporeal super-populations of the current populations. The number of entities in these two registry systems may be an order of magnitude larger than the current known population size. The registries can include "dead" entities or "chaff" generated by the complex multi-dimensional birthing process.

The BR's list of firms encompasses entities derived from many different indicators of economic entity activity in NZ. These varied sources can give conflicting or partial information concerning the presence of economic activity. These collection of administrative source files used to identify and birth entities we will call the **administrative registries**.

In the administrative registries, a significant duplication of entities and activity can occur. The entities birthed from the administrative registries into the full **entity registry** will represent a spectrum of firms. The spectrum spans the quality of the administrative information indicating whether the entity exists or not. At the top of the spectrum are firms with multiple confirmed indicators of existence from high quality sources and at the bottom are firms with only partial information from 1 low quality source.

All sources are not considered of equal quality or informative value concerning the presence of true activity. As such, rules must be created that define when the activity observed implies an actual birth to the registry. When maintaining the registry, there is a trade-off concerning the breadth of information included in the registry versus the cost of processing and maintenance this information. Typically, a small number of sources and source IDs are considered to be "core" indicators. New IDs from these core sources tend to always generate births. The internal IDs for these limited number of core sources are maintained and cleaned within the registry. Typically, these are maintenance processes ensure that the core IDs are consistent over time and space and unduplicated. If there are multiple core IDs on the registries there should exist an internally generated unique ID that spans the population of all the core sources. If hierarchical structures exist on the registry then multiple internal IDs may be required. Figure 8 summarizes these activities.

*Figure 8: Phase 1: The birthing and maintenance processes for generating a registry*

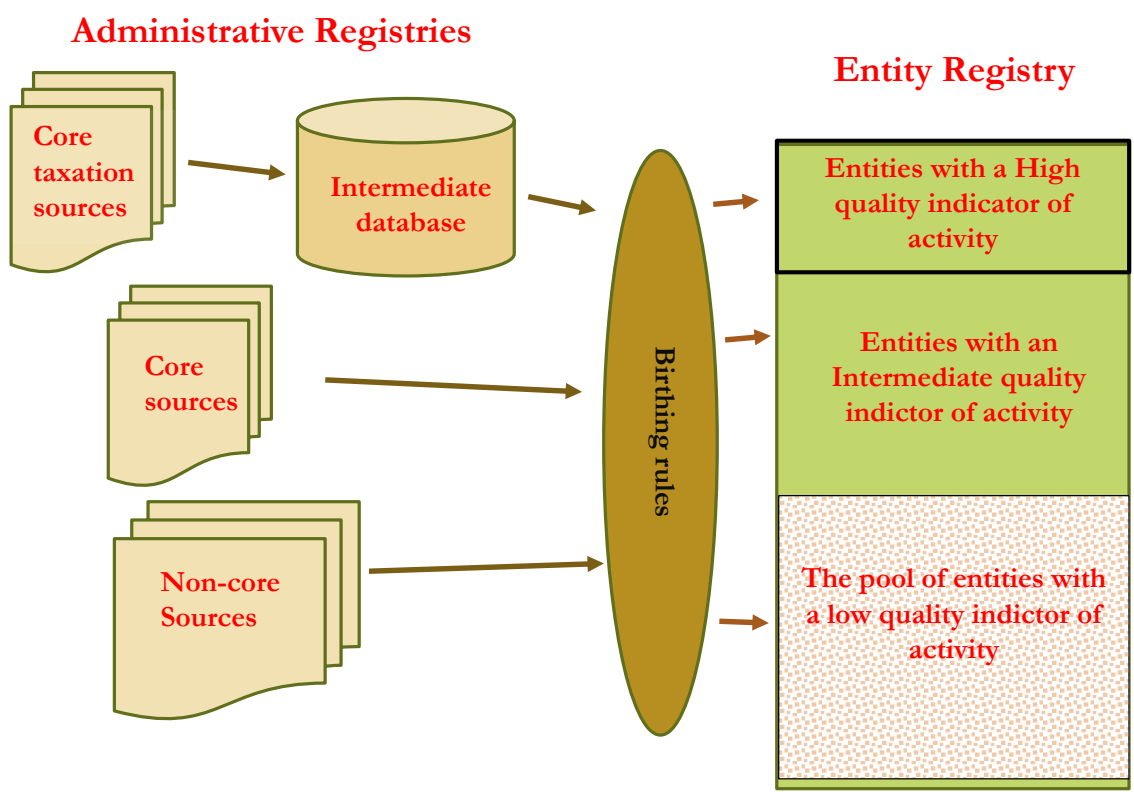

The above process is "under-the-hood" and not seen nor understood by external users of the register. The full registry is not useable in any real sense by outside users because it contains a collage of disparate populations and plethora of entities whose status is unknown. Frequently, the number of entities in the full register will be an order of magnitude larger than the estimated current full population. The fundamental process in Phase 1 is the "Birthing rules" shown in Figure 8. In the BR, these rules are typically set by a cooperative team from the BR, National Accounts, business surveys and methodology. External users of the registry cannot change or control these rules. Even BR-operations cannot change the rule without involving the full team. Presumably, the PR and LR would have similar teams tasked with defining the birthing rules for these registries.

The full registry system is rarely seen by external users. Instead, the users see the **statistical registries** (or target population frame[14]). The registry system presents statistical registries (or survey frames) to users by putting a filter over the full registry and each filter has viewing hole. Each filter changes what entities the user will see. Each user will not see the register in its entirety, they will only see what the holes allow them to see. This is the second phase of the registry maintenance system. Figure 9 below summarizes this process.

---

[14] A frame is the empirically derived list of the target population that is being estimated for. It is our best possible enumeration of our target population. In the design-based paradigm it would be our sampling frame and most practitioners think of a frame in the context of a sampling list but it is also our best possible estimate of the empirical core registry.

*Figure 9: Phase 2 creation of target population frame*

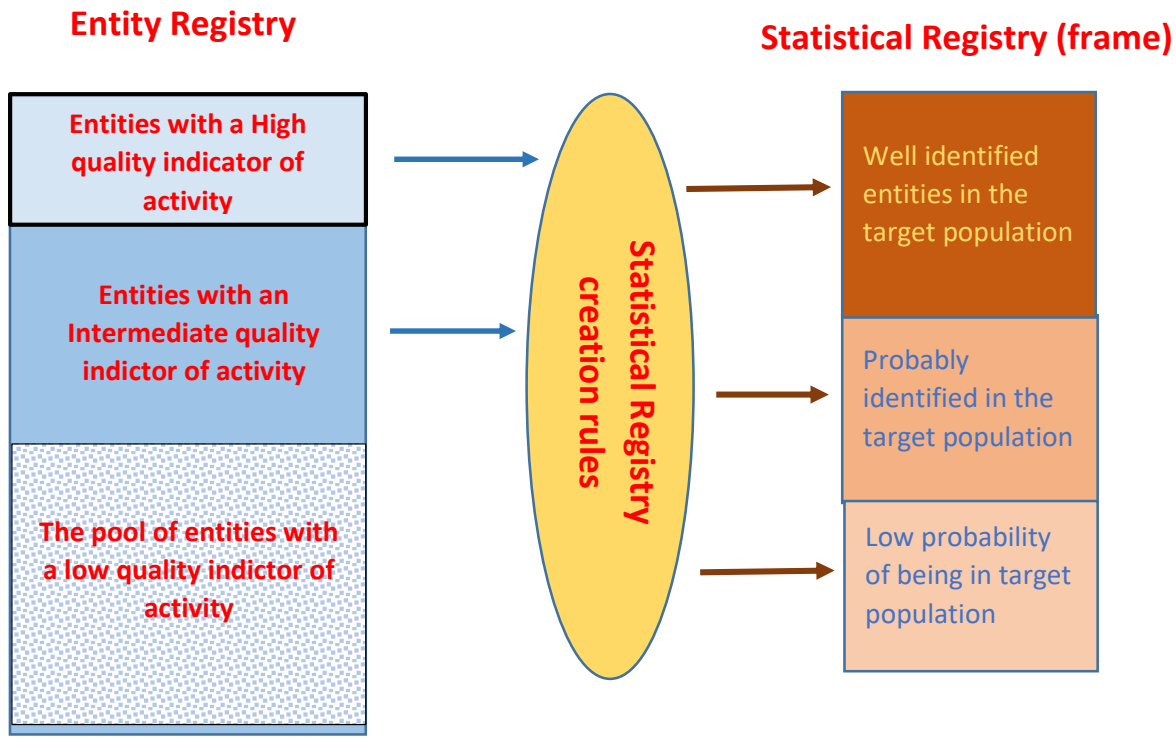

What the outside users sees are the statistical registries rather than the full registry. The core registries that users of our database will see are the statistical registries (BR, PR, LR) for the current best list of active firms or resident people or property holdings in NZ. Our core registries might be more correctly called core current target population frames which is a subset of the full registries. Traditionally, users of the BR blur the three phases of the BR system and they are all referred to as the register[15]. This blurring of terminology is relatively common when we talk about the Business Register (BR). Most users of the current business frame tend to refer to this frame as the Business Register yet this frame is just one mask through which we view the BR. We see the BR as presenting different faces (or masks) of the BR to different users.

The fundamental process in Phase 2 is the "Frame creation rules" shown in Figure 1. Again, in the BR, these rules are set by the same cooperative team from the BR, National Accounts, business surveys and methodology. External users of the registry cannot change or control these rules. Even BR-operations cannot change the rule without involving the full team. Presumably, the PR and LR would have similar teams tasked with defining the birthing rules for these registries.

One might think that external users might benefit from having control over designing the "birthing rules" and the "statistical registry creation rules" but this is not feasible. In general, these rules are arcane, complex and deal with multiple sources of information. The rules define which entities are current "active firms" or "resident persons" or "land holdings". The rules are capable of generating time slices for the target populations. Thus, the rules will generate consistent core registries for different time periods.

Typically, a central agency will generate a full business frame every major publication period (monthly, quarterly or annually) and they will take a snapshot of that publication frame & save it. Thus, one can always recreate the business frame that was used for publications. Meanwhile, updates keep flowing into the BR and influencing our view of these historical periods. In general, while revised historical frames could be created, the original snapshots are used instead. Presumably, this would be an appropriate strategy for the other core registries. These snapshots of the 3 core registries will bound are estimates and allow use to calibrate for missing responses.

---

[15] The PR register system at Stats NZ is referred to as the **"spine"** and like the BR it is a complex system with multiple different registries that are all referred to as the spine.

The Statistical Registry will be used to calibrate the estimates and hopefully make the results representative (unbiased) of the target population. Below is the second phase of the estimation process from Figure 6 where we estimate a health versus income relationship. Figure 9 shows where our core registry might fit into the process.

*Figure 10: Phase 3: Using the Core Registry to Calibrate Estimates*

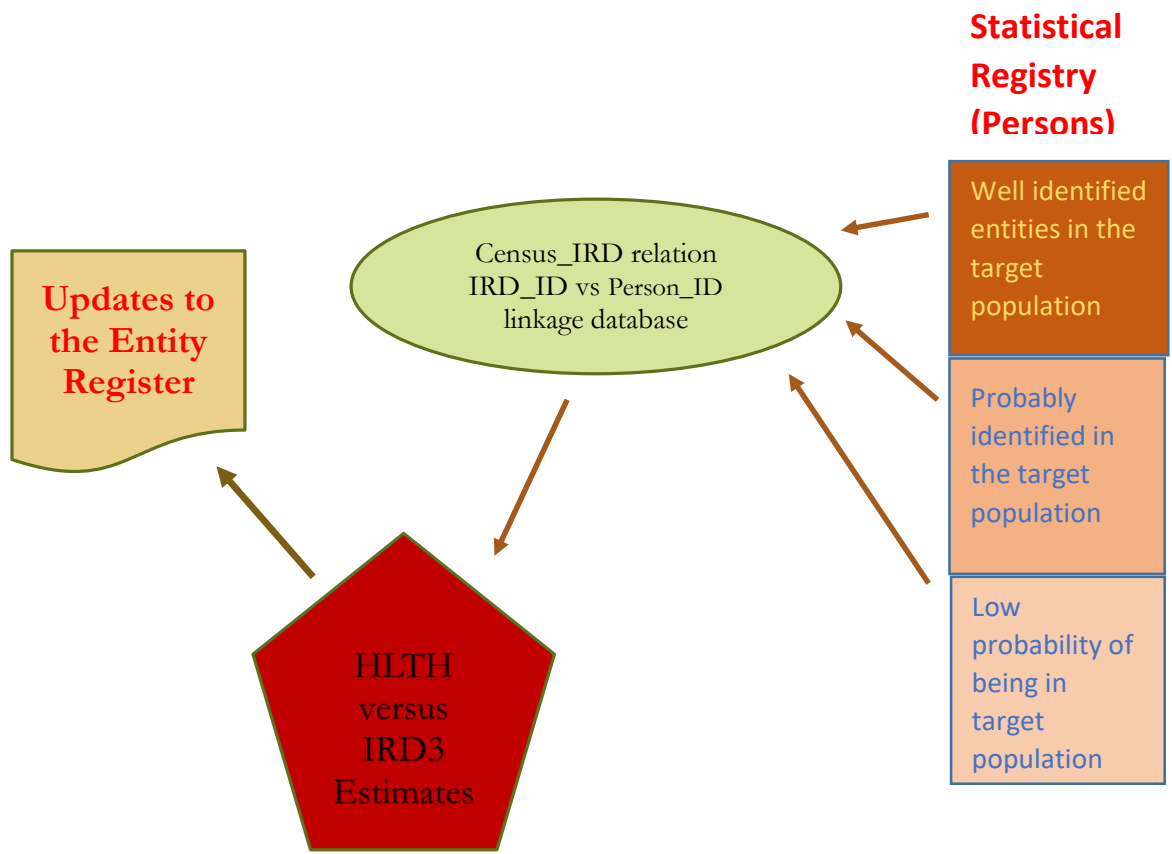

Figure 9 illustrates that a feed-back loop may exist between these estimation process and the registry processing system. In the context of the BR, business survey responses can lead to name, address, industry classification, etc. updates and these are sent to the registry maintenance staff for integration into the latest BR. These feed-back loops are important for keeping our PR core-registry as up-to-date and accurate as possible. As the CSAs get more proficient at generating these administrative estimates, they will provide the PR with an ongoing grooming function. Figure 10 integrates all three phases into one integrated registry maintenance system.

Figure 11: Putting it all together: The Registry System

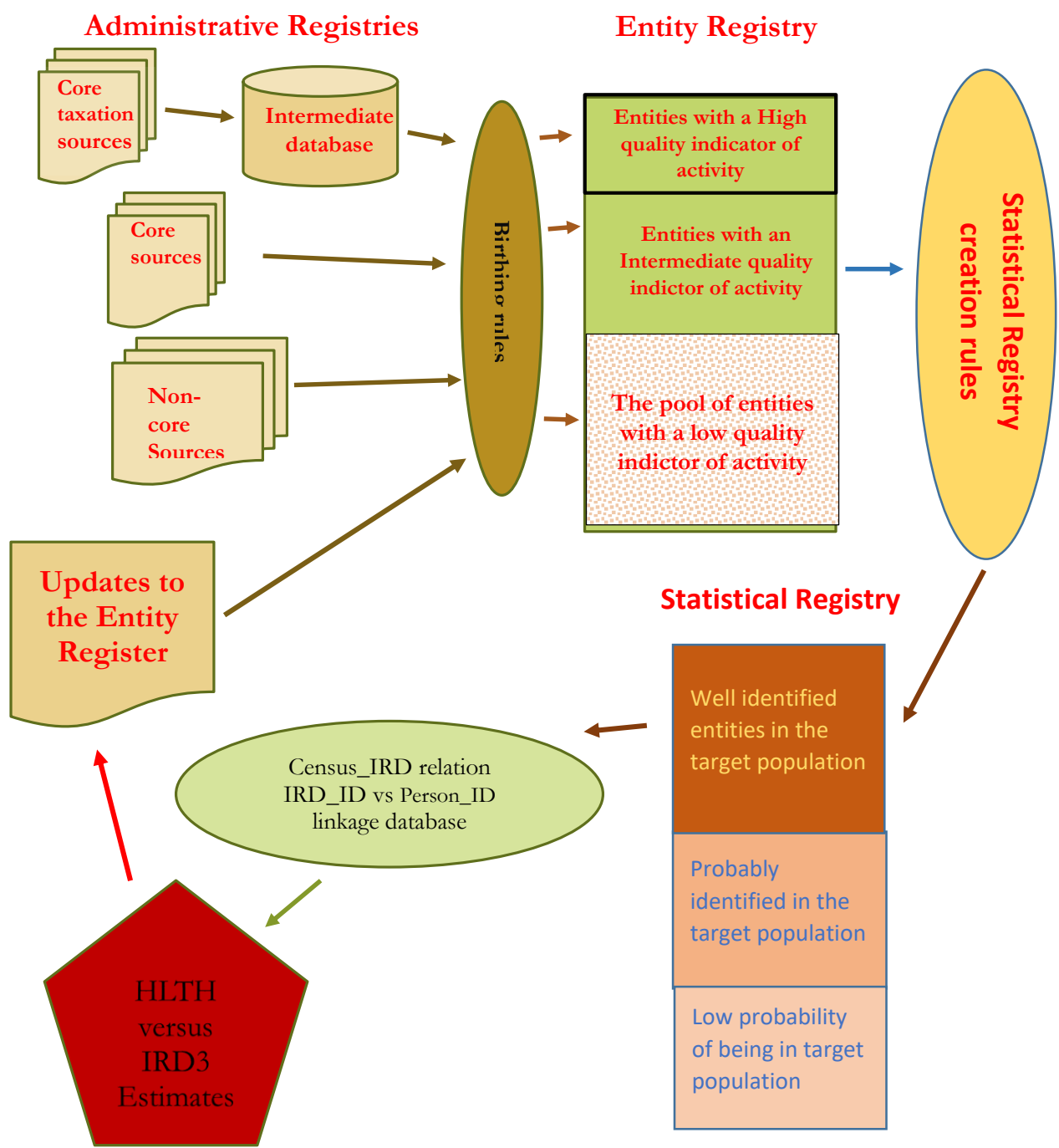

At Statistics New Zealand, the full **schema** covering the core registries, the various source files and the linkage functions is referred to as the **Integrated Database Infrastructure** (IDI). The key **administrative registries** are the administrative birth, visa and income tax registries. The **entity register** is referred to as the population **spine** and the **statistical registry** is referred to as the **estimated residence population** (ERP) registry. Users of the database generally do not directly interact with the administrative registries nor the spine (the entity register) but instead interact with the statistical registries and the source data files for information linked in their project.

# 6 META-DATA IS THE GATEWAY TO THE EVOLUTIONARY SCHEMA

The evolutionary database schema is a conceptual framework rather than a detailed design plan. The evolutionary schema implies that that source datasets, registries, linkage relationships and estimation strategies will constantly evolve. Underlying database paradigms may evolve from traditional relational models to newer structures such as schematic models. New data years, new sources, new algorithms, newly discovered relationships and new access modules can fundamentally change the quality of the evolutionary database and the manner in which it is used. In such an ever-changing schema instabilities, inconsistencies and confusion about the current state of the evolutionary database can become insurmountable problems if methods are not implemented to control these issue. Meta-data can be the fundamental quality control mechanism in the evolution schema.

Researches will often approach the evolutionary database with an ambiguous research objective (or questions). While these research questions may appear to be wide-ranging or imprecise, they often have very restrictive underlying constraints that will impact on estimation: such as requiring specific data years and/or subpopulations and/or source datasets and/or relationships being estimated. Researchers will want to know if the available sources/relationships/registries/linkage functions adhere to these constraints. Thus, associated with every object/function in the evolution schema should be a meta-data descriptor.

## 6.1 STANDARD DESCRIPTORS OF DATABASE ACCESS, FILES AND FIELDS

The most basic and most requested meta-data is an explanation of how to access the data and how it is structured. Users wish to get on with using the data. They require database access protocols; file names and locations; field names, formats and brief descriptions; and source providers for the various data sources. Users tend to fixate on fully opening the doors to the data rather than understanding the data's quality, usefulness for their question and its limitations. Unfortunately, sometimes the meta-data associated with CSA databases caters exclusively to maximizing the ease of access for users.

While this access information is vitally important to users, it presents dangers if it is not balanced with information related to the quality and limitations of the data. Without some discussion of the populations covered by each file, you are encouraging users to apply their analysis to inappropriate sub-populations. Or perhaps the user is linking two sources with slightly different entities (family versus head of household) and this creates misleading relationships. The authors believe that the first priority of meta-data should be the presentation of quality information to the database user. This is a tall challenge yet the database owners must recognize that fixating solely on the most requested and easy to create meta-data has a dark side. In these circumstances, the database owner is encouraging users to miss-use the data. We believe CSAs must accept responsibility for the data they provide and that means providing quality information in the database meta-data.

In the following sections, we first discuss the potential dimensions of quality to include in the meta-data and in the next section we propose a semi-automated strategy for measuring one particular measure of quality: coverage.

## 6.2 META-DATA AND THE SIX DIMENSIONS OF QUALITY

Quality is a multi-dimensional issue in survey design and it involves both measurable and non-measurable-judgemental indicators of quality. Most CSAs propose evaluating survey quality in 6 dimensions: relevance, timeliness (or alternately punctuality), accuracy, accessibility (or alternately clarity), interpretability (or alternately comparability), and coherence. The 6 dimensions can overlap occasionally but mostly they are distinct dimensions of quality. Omitting any dimension could cause a methodologist to over or under estimate an important aspect of quality.

For researchers, analysts and other users of the database, there will be critical questions whose answers should be recorded in the meta-data associated with each dataset. This information will constrain the researchers' potential analysis and results. When a source dataset is inserted into the evolutionary database, methodologists should create meta-data associated with this dataset that relates to the 6 dimensions of quality. They need to carefully think about all 6 dimensions and how they are relevant to the source dataset. They might consider questions such as:

**Relevance**: What type of entities and events are covered by the datasets? Are there any special sub-populations not covered by the datasets? Are there any special sub-populations covered by the datasets that may not be in a typical target population? What data fields are contained within the source datasets? What are the definitions of the fields stored in the database? Are there any special non-standard definitions used in creating the fields in the datasets?

**Timeliness**: What data periods are covered by the datasets? Is the data received on a fixed periodicity? What is the time lag between the observation date/time and the load date when the data is loaded into the evolutionary database? Are there any data periods where data is missing? Were there any special time observations where an unusual event might have affected the data? Does the quality or the types of the fields loaded change over time?

**Accuracy**: How accurately is time recorded in the source datasets? How prevalent is missing datasets or missing timeliness or missing events or missing data fields? Does the dataset contain identifies that uniquely define entities? Can events or entities be duplicated in the dataset? Can entities be linked to multiple identifiers? Can multiple entities be linked to the same identifier? Are there any indicators of the quality of each data field perhaps a measure of miss-coding for a field?

**Accessibility**: Are the datasets within the evolutionary data base standardized for ease of access? Are some of the datasets stored in older hard to access formats or locations? Are there special access modules written to improve access by non-experts? Are some of the databases, entities, events, etc. not accessible due to confidentiality, privacy or quality issues?

**Interpretability**: What are the factual definitions of the various fields in the datasets? Are the definition of the fields, events or entities non-standard in any way?   (As an example, in many datasets there is a field called "income" yet this particular concept can cover many radically different definitions that can affect the variable's size by orders of magnitude.) Is the convenience sample or the "is-what-it-is" dataset representative of the target population? Can the estimates be interpreted in a clear and straight forward manner?

**Coherence**: Does the concept of entity change across time, space, event or timelines? Is time recorded in a consistent manner that allows proper ordering of events and integrating of datasets? Are common variables consistently defined and available across time, space, event and timelines? Does the coverage of the target population change in the "is-what-it-is" dataset change across time, space, event or timelines? Are we sure that the relationships between entities, events and datasets are stable across time and space?

These are just some of the questions that that we might attempt to address and answer in the meta-data-associated with each dataset. Measuring quality is a complex issue and almost never quantifiable into a single dimension or number. Collecting and recording the answers to the above questions can represent a significant investment in ongoing resources.

In practice, few CSAs invest the resources necessary to address all these questions and instead tend to confine meta-data to information required to access and use the data. These abbreviated meta-data records tend to define little else besides dataset locations, dataset format, fields available and formats of those fields. It is a worst case scenario.  Users are given maximum accessibility bereft of any quality information. In the next section of the paper, we will outline an automated strategy for giving users a minimal set of statistics relating to quality.

## 6.3   THE BASE REGISTRIES AND THEIR IMPORTANCE

In the paradigm that we present, there are 4 fundamental or core dimensions: time, space, firms and people. Like in science, time is a transcendent dimension that cannot be enumerated nor converted into a registry. It always exists and every piece of information in the evolutionary database must be time stamped but because it is a continuous variable with a single direction, infinite past and non-existent future it must be treated differently than the space, people and firm dimensions. Space, people & firms are enumerable sets of entities and as such we can create registries that enumerate these 3 types of entities at a specific point in time. These 3 registries are our "**base registries**" within our evolutionary schema. They anchor our estimates and shine light on the quality of our estimates (Figure 11).

*Figure 11: The 3 base registries illuminate the quality of the convenience sample*

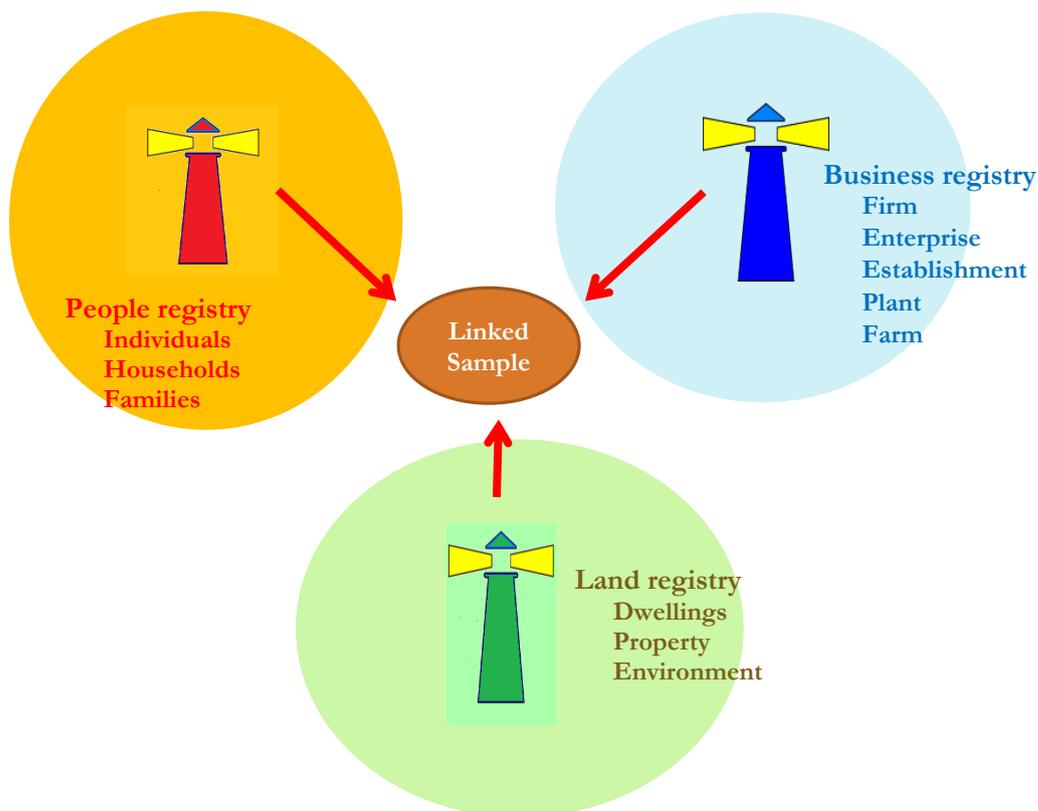

Our "**base**" quality measures will be defined as the coverage ratios of the 3 base registries for the convenience sample.  While good coverage of the 3 base registries will not guaranteed the quality of the estimates, it will tend to be a necessary prerequisite. If the convenience sample is representative geographically, by people demographics and by firm industries it does give one a much greater faith that the estimates will be representative of the target population.

Initially, we might only have linkage relationships between one of our base registries and the convenience sample and thus we would only able to generate coverage ratios for 1 dimension. Then as we bootstrapped the evolutionary database, we might eventually develop linkage relationships for the other 2 base registries and improve the quality measures to cover all 3 dimensions. We see this process as a distributed evolutionary method that solves problems locally and slowly bootstrapping the quality of the overall system.

We see representativeness as the Achilles heel of integrating disparate "it-is-what-it-is" datasets and we see base registries and their evolution as the answer to that problem. It should be noted that while we discuss the schema mainly from a design-consistent framework, we could have presented the same paradigm from a model or Bayesian framework.

# 7 TOTAL SURVEY ERROR MEASUREMENT IN THE EVOLUTIONARY SCHEMA

Total Survey Error (TSE) can rarely be reduced to one number, instead it is a structured methodological approach for reviewing and compiling sources of error in a survey. Errors can arise at each classical survey processing step: frame creation, sample design, questionnaire design, questionnaire distribution, collection, editing, follow-up, imputation and estimation. Each survey processing step can introduce errors in potentially different dimensions of error. The TSE paradigm treats each aspect or dimension of the survey processing system as part of a collage that defines the overall measure of quality. At each processing step, quality measurables are collected and assembled into an overall package that allows the survey designer to understand where errors occur and give them some sense of their overall impact on the TSE. Traditionally the public and media fixate exclusively on errors related to sample design (the infamous 19 out of 20 media statements or 95% confidence interval) while survey methodologists recognize that questionnaire design, collection and non-response adjustment are the major contributors to error in many surveys. TSE is a structured approach to cataloguing and measuring the errors that arise in each of the classical survey processing steps. TSE also addresses dimensions of suitability or errors that are not well defined, are fuzzy and are not easily quantified, such as, timeliness, relevance, and consistency. TSE challenges us to view survey errors in a structured and holistic manner.

While the classical survey processing steps may not be applicable in a data integration paradigm, we will be following a similar structured approach that breaks down the overall estimation process into self-contained sub-processes. Measurables and indicators will be suggested within each sub-process. Using our example of investigating the causal relationship between wealth and health we will proceed to develop a TSE framework for Figures 5 and 6.

## 7.1 THE PROCESSING STEPS AND THE DIMENSIONS OF TSE

Classical survey processing theory breaks down the overall survey processing into standardized widely accepted sub-processes, such as: questionnaire design, frame creation, sample design, questionnaire dissemination, questionnaire collection, editing, follow-up, imputation and estimation. In the TSE paradigm each sub-process is examined independently from a quality measurement perspective. Thus we have a vector of 9 quality measureables, one for each sub-process. In the TSE paradigm, quality has 6 major dimensions: relevance, timeliness (or alternately punctuality), accessibility, interpretability (or alternately clarity), coherence (or alternately comparability), and accuracy. In the TSE paradigm quality is not one number but rather a matrix of 9x6 quality assessments. In the TSE paradigm a survey methodologist is required to view quality in a multi-dimensional yet holistic manner.

The TSE paradigm can be applied to integrated dataset but the names associated with the black boxes describing containing the sub-processes must be renamed.

*Figure 12: sub-process in the creation of an integrated file from 2 source files*

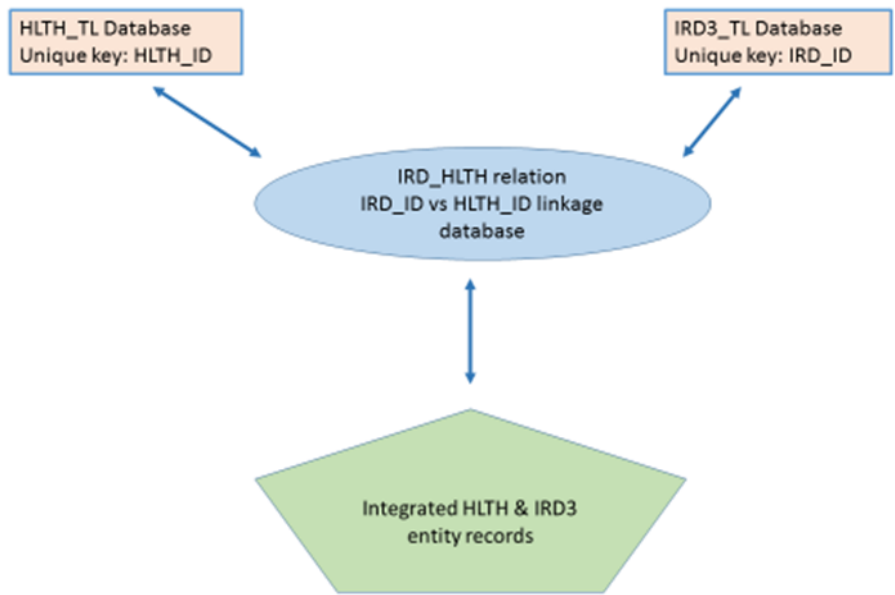

Figure 12 illustrates the first phase of the integration process given in Figure 5 where we are integrating health and wealth data. In our paradigm, we are always doing direct linkages of entity IDs occurring in two source datasets. We are never integrating more than two distinct sets of entity IDs. When we wish to integrate more than 2 sources we will step-through the data. This is illustrated in Figure 5 where in the first phase we are integrating health and wealth information using the HLTH_ID and IRD_ID linkage relation and then we integrate the this output with the person registry in phase 2 by using the IRD_ID and Person_ID linkage function to get the final integrated estimation file. Within each phase we have three distinct sub-processes that could lead to error creation. Figure 12 illustrates the first phase of our example and the three sub-process are the creation of source files HLTH_TL and IRD3_TL (beige rectangles) and the creation and use of the IRD_HLTH linkage relation (blue oval). Thus, in each integration phase we will gather quality measurables applicable to each sub-process. If as in Figure 5, we have multiple integration phases we will just continue the construction of our collage of quality information. Our quality assessment report will be a compilation of the error arising in the source datasets, the linkage functions, the registries, and finally the estimation formula.

## 7.2 ERRORS ARISING FROM SOURCE DATASETS

In our schema the source datasets are the beige rectangles in Figures 4, 5 and 6. Source data sets will be administrative, survey and census datasets containing observed variables that we wish to model or inter-relate. Often the incoming quality and content of these sources will be beyond the control of the CSAs. In the following section quality measurable will be suggested for these source datasets. Note that timeline concept introduces a separate potential manner in which to view quality in the schema. We will continue to use our examples of the three source datasets defined in sections 2 and 3.

### 7.2.1 Basic information about the event data sources

Any researcher or analyst using the evolutionary schema needs to understand the strengths and weaknesses of the event source data before beginning their analysis. Thus the evolutionary schema should provide basic information that might impact on the quality of the analysis. Note here we are For instance we might address some of the following questions.

- Which agency collects the information?
- Is the data collected for a legal purpose and if so to what purpose?
- For what type of entity is the response appropriate?
- What time period does the data cover?
- What is the frequency of collection?
- What variables are collected?
- What is the definition of the variables?
- Is any entity exempt from reporting?
- Are there known weaknesses associated with the data?
- Have there been any discontinuities in the collection or definitions of the variables?
- Is there any missing data in the dataset and how prevalent are these missing values?
- Is any of the data values imputed?
- Do we have any information on miscoding errors, etc.?

Over time, other general information about the source could be collected and stored but there should be a common set of questions crossing all data sources. In general, the data source files will be collected by non-CSA agencies for non-CSA purposes and control of the quality and content of the source files will be outside the CSA's purview. For this reason, it is important to identify the true weaknesses and strength of the source datasets.

Administrative datasets, like tax filing, often appear to collect the same information from a range of disjoint sub-populations. Yet, common variables used across all the data sources, such as "revenues", can have significantly different definitions and meaning in the different tax-filing datasets. In addition, the reporting entity in the various datasets can be subtly different and the various datasets can be disjoint or partially overlapping. Over and under coverage can vary markedly across the datasets. All these issues present serious challenges for any researcher that wishes to produce consistent estimates for the full target population. Proper documentation of the incoming quality of the source datasets is a critical element in measuring TSE in the Evolutionary schema.

For our IRD3, HLTH and EDUC databases we should specify what data years are covered, give reference to any legal act bearing on the data and the data source and provider. We should detail how the reporting datasets might have evolved over the timelines. Perhaps in both the education and health data new tests were developed and obsolete tests were abandoned on specific dates. In our IRD3 dataset perhaps the law was changed requiring the reporting of new information. Perhaps in health or education, the IDs changed over time because of the introduction of a new system. If the coverage of the New Zealand resident population changed in any of the 3 data sources that needs to be noted.

Each new data source introduced into the evolutionary schema needs to be researched to uncover as much information as possible that might impact on the quality of estimates derived from this data. If the stored data changes over time because of non-causal effects the researchers need to know this fact. This is key requirement in building the TSE picture for any integrated dataset analysis that will be used in cause and effect analysis.

### 7.2.2 Coverage statistics

In the Evolutionary Schema, population coverage is a critical concept because it is expected that most data sources will have biases in their coverage of the target population. Every source dataset should be related to its coverage of a frame or registry, preferable either the business register or the person registry or the dwelling registry. The population coverage should be as devolved as possible. Thus, for person entity data sources one should provide coverage by sex, age, geography and as many other demographic characteristics as possible. As an example, let us consider our 3 example data sources.

One might suspect that the IRD3 data undercovers the young, the old and perhaps females. One might also suspect that coverage will be affected by demographics such as education, health, immigration status, involvement with the justice system and wealth. One would expect that education data might only cover relatively recent time periods and persons who were resident in NZ during their education. Thus one might suspect the data under-covered older persons and immigrants. Many studies have shown that health expenditures are concentrated in seniors and young children. In addition, health expenditures are significantly affected by wealth. Thus one might expect that the health data under-covered middle-aged and poor people. If in fact, these assumptions are borne out by the data, users need to be made aware of these facts. The basic coverage statistics that we propose generating will be an important factor in alerting users about potential problems.

### 7.2.3     Stability over time

What we are proposing here is analysing changes within event timelines. Timelines are collections of events for a specific entity from a common data source. For each entity's timeline we could record statistics concern how often a field changes or is missing. Then aggregate (domain) estimates of the average changes or proportion of missing value could be automatically generate and placed in the meta-data. Grouping events into timelines gives us an extra dimension of quality to measure.

### 7.2.4     Evolution will develop new measures of quality

As new registries, sources, linkage functions, timelines and events get added to the database, users will develop new insights about the data and how to measure quality better. We will discover new algorithms to calculate these measures and place them into the meta-data.

## 7.3   ERRORS ARISING FROM THE LINKAGE FUNCTION

If we are fortunate enough to have one common entity ID that is universally coded across all the administrative files, then the linkages functions would be theoretically error-free. Unfortunately, except for in a few Scandinavian countries, a common ID for persons does not exist within most countries. I many countries there is a significant political movement opposing the creation of such an ID because it is perceived as an invasion of privacy. Many countries (like New Zealand) have explicit laws that limit the usability of IDs that identify specific persons.

While privacy concerns are likely to be less relevant for firm and land entities, universal IDs for these entities are rare. IDs for these entities have tended to be created to suit local needs and thus a profusion of inconsistent IDs exist. In this case, there is no economic incentive to create and pay for the maintenance of a universal ID. It is more efficient and effective to invent a new set of IDs for each usage. Why should a municipality pay for the installation of a complex and expensive land ID system spanning a country when they can do the same thing with a ledger and one part-time employee? Integrating has a cost and the question is who will pay this cost? Furthermore, the municipality may be able to embed local information into the ID and this might offer an ancillary benefit to the local municipality. They might lose this extra benefit in a generic universal ID system.

Without these universal IDs, linkages across administrative data must be based upon matching common information contained on the two files. Fields such as name, address, age, sex, etc. are used to identify the best links. The difficulty is much of this information in these "linkage" fields is self-coded with little thought given to standardizing or editing the way information is encoded. In this case, the linkage function uses fuzzy, imprecise and possibly conflicting logic. Empirical fuzzy concepts such matches, probable matches, probable non-matches and non-matches get defined. Issues such as false positive matches and false negative non-matches arise. In addition, there is the problem of identifying and eliminating duplications when the knowledge base is incomplete.

This is a complex specialty field and a number of strategies and algorithms have been developed to address this issue. Some of the methodologies developed calculate pseudo-probabilities and can quantify the occurrence of some types of errors. Where these measures are available they should be included in the meta-data.

Our schema does offer a naïve but straight forward method for measuring one dimension of quality (representativeness). Coverage ratios relating the linked subset to the 3 core registries can be calculated. The coverage ratios for the intersection sets defined by the linkage function can be calculated by various demographic, geographic and economic domains. This could be auto-generated and stored in the micro-data.

Lastly, the linkage function's algorithm should be documented together with a brief summary of its strengths and weaknesses.

### 7.4 ERRORS ARISING FROM REGISTRIES AND FRAMES

As noted in Section 4 of this paper, registries are complex multi-file collections of entities with auto-regressive feed-back loops. We often refer to this collective group of files as if they are one file. Users tend to only see the "faces" of the Entity Register and they equate the word "register" with a particular "statistical register". That represents their target population. Errors and quality issues can arise in all 3 phase of the register.

#### 7.4.1 Administrative Registries

Errors can arise within the Administrative Registries used to create the Entity Register. The primary areas of concern are over or under-coverage of the Entities in the Administrative Registers plus hard to detect duplications of entities. Of lessor concern but still an important source of errors is miss-coding, missing fields, and non-standardization of field content in the Administrative registries.

Under/over-coverage in Administrative Registries can arise at various hierarchical levels and in subtle ways. It can occur at the entity or at higher aggregate level. In the case of the administrative health data we discussed in Sections 1-3, it is possible we only have health data for a limited number of the total hospitals or regions. In addition, if we are recording events related to hospital visits, perhaps there are individuals who have never visited a hospital. Or perhaps we only have administrative data from private health insurance sources. Thus over/under coverage can be created by a number of different mechanisms. In some cases, we may be able to quantify the level of over/under-coverage but in other cases it may be impossible to estimate. As an example, let use consider the case just mentioned: only some hospitals are covered.

In this case, we could construct a registry of hospitals. With this known list of hospitals and some ancillary information we might be able to estimate under-coverage effects related to the missing reporting hospitals. Enumerating and quantifying all the sources of error arising from the Administrative Registries is extremely resource intensive and an evolving TSE system that incrementally adds new quality related information can help spread the work over a long-time period.

## 8 SUMMARY

We set out to address what we felt were the key issues in using administrative data: estimation and assurances of quality. In a two year-long discussion amongst ourselves and external methodologists we explored the numerous pitfalls one encounters when using administrative data and we discussed several strategies that needed to be a part of any statistical system using administrative data. While we quickly realised that representativeness was the Achilles heel of administrative data, we were also strongly influenced by Zhang's (Zhang, 2012) recent paper calling for a new conceptual paradigm when dealing with administrative data. Thus, right from the beginning of our discussions we attempted to tackle the problem in a holistic manner and we attempted to create a full conceptual paradigm for dealing with administrative data. Below is a summary of our major finding and an outline of the main features of our schema.

### Non-representativeness of administrative data

The key weakness of the administrative data that is currently available to Statistics NZ is its under-covers various sections of the current population. Coverage is never 100% in any administrative data source and in almost all cases, significant portions of the current population are under-covered. This is a systemic problem that is considerably worsened by cross-linking multiple administrative datasets. If one links a dataset that under-covers women with another that under-covers children, you get a linked dataset that under-covers both women & children. This is the Achilles heel of administrative data & if we cannot address this issue we will never be able to widely use administrative data.

How can we defend ourselves from criticisms of bias caused by under-covering the under privileged or the rare populations? Without a methodology to measure under-coverage & correct its effects how do we maintain our credibility?

### Correction with registers & frames

Our solution is to create the 3 lighthouse registers & use them to measure under-coverage in various domains/stratums/classes. Then we use these registers to create calibrations (design-based designs) or models (model-based designs) or Bayesian priors (Bayesian designs) to correct for under-coverage. This strategy is both naive and simple and probably will not correct for all the biases created by the administrative datasets but it is a first step. We are following the historical path of development of corrections for census non-response. We consider this a first step in the correction for non-response bias. As we gather more expertise we will develop more mature and complex methodologies.

### Evolutionary (system grows in every sense over time)

We see the paradigm as an evolutionary system in every sense. New data source will evolve and old ones will disappear. New data points will be added (both in time and cross-sectionally). New database designs will be incorporated and new estimation and linkage algorithms will be constantly developed. New methodologies will constantly be under development and being evaluated. Our testing of our estimation and error measurement techniques will grow as we acquire greater understanding of the systems.

### Distributed and collaborative system

Administrative data spans wide arrears of knowledge, subject-matter areas, geography and time. In addition, the final design should incorporate ongoing development of complex statistical and IT methodologies. No one group can centrally do these tasks. The tasks and data sources must be delegated across various teams with varying backgrounds and expertise. Of course a central design and control structure would be created to oversee these teams.

### Evolutionary Convergence

When we create our registers, data sources, methodologies, etc. there must be a path of convergence towards a final optimal system. Ideally each new evolutionary step will incorporate all previous information gathered. Consider the BR. In general, most of the information in the BR is tombstone information that rarely changes over time. Births and deaths are a small percentage of the population, only a small percent of the address or names change each period, etc. The BR staff focuses on changes rather than the full population. This is also the manner in which the census address list is maintained. Similarly, our evolutionary system would be built along paths that evolve towards better quality and optimality. The system must focus on changes in the population rather than all the population.

Feedback loops are important in this system. Users especially internally must be able to feed corrections that they have identified in the registers, algorithms, etc. back into the system. This is way the BR and the census address works. This helps us catch evolutions in our data sources and it leads to a convergence path.

## Timelines (cause & effect)

There is one message we are hearing continuously from external researchers. They want to do cause and effect studies or longitudinal studies. Our data need to have a natural time structure built in from the beginning. We see each administrative data transaction as an event that occurs at a specific time. We propose grouping and time ordering these into timelines of events that occurred for a common entity. With each data source an implicit timeline database would be created for each entity.

Viewing the data in this manner not only allows for cause & effect studies but it opens new possibilities for editing, imputation, linking, etc.

## Total Survey Error

We propose a simple and naïve strategy for creating measurement tools for estimating Total Survey Error (TSE). We propose calculating in each stratum/class/domain defined by the 3 lighthouse, the coverage ratio of that data sources versus the estimated light house population. When cross linkages are done the stratum coverage ratios should be calculated for the linked dataset. This TSE information should be stored in the metadata.

## Metadata

The metadata will store all these coverage ratios plus estimates of the true current stratum sizes. When a researcher comes along & uses the datasets they will automatically get a profile of the coverage ratios and stratum sizes for all the data sources they will be using files plus they will get the same information for their linked dataset.

## Survey design

The way we have structured the paradigm, it naturally allows us to re-use standard historical design practices from any of the 3 major design areas. Of course we know that administrative data is not probabilistically generated and these datasets often contain complex generating mechanisms with unknown deterministic generating functions that create statistical biases. Yet, we are naively assuming that within a design stratum the selection method for generating administrative records is random. This is a naïve assumption but we consider it a first step in developing a new paradigm. We will use standard off the shelf survey methods as our first initial step.

# 9 APPENDIX A: BUILDING A POPULATION REGISTER

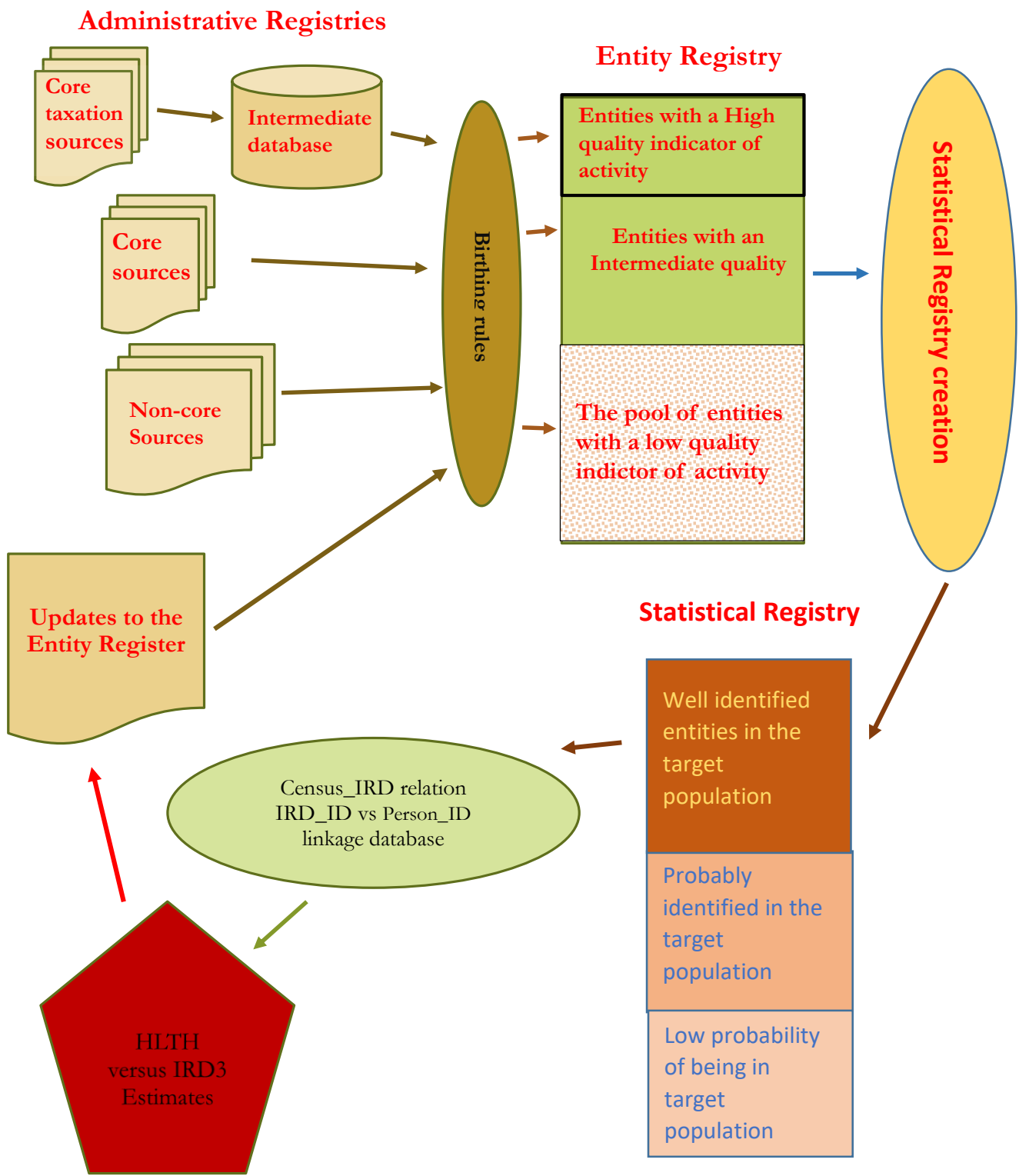

# 10 APPENDIX B: GENERATING A VIRTUAL ID FOR A POPULATION DERIVED FROM MULTIPLE SOURCES

## 10.1 FIRST STEP – ASSIGNING IDS TO ALL INCOMING SOURCES

### 10.1.1 The Virtual ID generator

The generator will create unique 15 digit IDs where the first 14 digits are a sequence ID that commences at 10000000000000. (Starting at this number ensures that format conversions between number and character formats are unaffected.) The fifteenth digit will be a modulo 10 check digit. The check digit will help limit manual coding errors. The generator will be very simple. It will contain one input register and two output registers. Input will be "seed" and output will be "$seed + 1$" and "$10 * seed + modulo$". Once pulled from the ID generator, this identifier must never be pulled again. This generator will provide unique sequential virtual identifiers (SV_ID) to all source data sets that are used to define the population. We are not using this generator on all data sources in our database. We are only using it in combination with data source used to define the target population.

In the case of the IDI, there are currently 3 core source data sets: IRD, immigration and birth administrative records. A fourth source is required to allow for feedback and manual updates. In general, one could have as many sources as one wished but the more there are the more complex the system becomes. Figure 1 below shows how this might work.

Figure 1

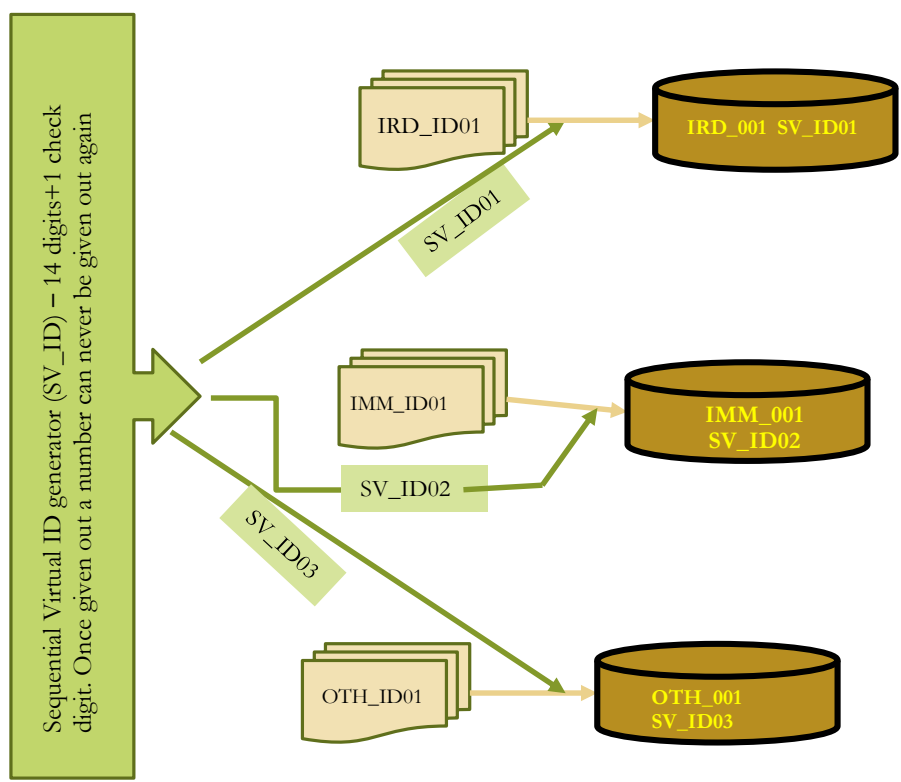

### 10.1.2 Loading the incoming administrative registers

We will the call the green disks in Figure 1, administrative registers. There will be one per administrative source used in the building of the population register and the primary lookup key in this administrative register will be the unique identifier supplied by the source organization. When a new transaction is received from the administrative source (say IRD), we will look up the primary key and if it is already in the register then we will update the information associated with this key. If it does not exist we will create a new record in the administrative register and assign it the incoming primary key plus we will ask the Sequential Virtual ID generator for a new identifier for this record.

Note that this process will create unique virtual entity identifiers that spans all of our potential source datasets.

### 10.1.3 Handling multiple identifiers assigned to the same person

In this case we have an individual who has been assigned multiple identifiers that are causing implicit duplicate entities in one of our administrative registers. Initially we will assume that all incoming transactions from a data source (say IRD) will have unique source identifiers and we will birth a new entity on the administrative register whenever a new primary key is received.

If we have a source or process that identifies duplicated primary keys then we will create another a secondary identifier which we will call an alias identifier. In this field we input the oldest or smallest sequential virtual identifier that was observed used for this entity. Figure 2 shows the case where we have 3 separate IRD_ID that have been assigned to the same individual.

Figure 2

|  | **IRD Identifier** | **Assigned SV_ID** | **Alias Identifier ALIAS_ID** |
|---|---|---|---|
| **Transaction 1** | IRD_001 | SV_ID01 | SV_ID01 |
| **Transaction 2** | IRD_002 | SV_ID02 | SV_ID01 |
| **Transaction 3** | IRD_003 | SV_ID03 | SV_ID01 |

Unique entities in our administrative register will be identified by having SV_ID=ALIAS_ID. Note that we must not delete the duplicates because these records define our paths back to the original source files. Our output data set that we will be feeding to the population register will contain all those records where SV_ID=ALIAS_ID.

### 10.1.4 Handing duplicate identifiers assigned to different entities

In this case, the same source identifier has been assigned to multiple individuals. It could be a situation where the incoming data source is generated by multiple separate institutions who each have independent algorithms for defining the primary key identifier. Because the generating mechanisms are independent, random duplications might occur. Another possibility is every year, the source organization might regenerate new source identifiers.

How we "fix" this issue is by finding a way to make the source identifiers unique across institutions and/or years. Thus we might define a new incoming identifier that appended an institution code and/or date code to the original identifies. The worst case scenario is that we might decide that the source identifiers are not truly keys in any real sense and give every incoming transaction a unique SV_ID. It is because of this possibility that I suggest that SV_ID be at least 14 digits. We might need a lot of overhead in the assignment of our SV_ID.

WORKING PAPER## 10.2 SECOND STEP – INITIALIZING THE POPULATION REGISTER

In survey methodology terminology a register is a container holding multiple disparate populations of related entities. The Business Register (BR) is an example. The BR encompasses the population of business firms that were active at any time during a specific time period. As an example, the BR may cover any firm that was active from 1950 until the current day. Thus implicitly the BR can be used to define the active population for any specific time point. Thus we could extract from the BR the business population for year 1990 or the business population for year 2000. Each of the entities in the various sub-populations must have a unique sequential virtual identifier (SV_ID) drawn from a single Virtual ID generator. Thus our SV_ID will be consistently defined for all input population indicators of activity and for all possible output populations.

So let us start to build our person registry from Figure 1. Initially there will be no attempt to link entities in the 3 sources nor remove duplicates. What I am proposing may appear to be overkill with needless duplication but all the records and fields are required if we are to have maximum flexibility plus the ability to constantly edit and refine our information in a manner that allows convergence and guarantees that update feedback loops always work. So initially, our register will contain every entity in the 3 source datasets where (SV_ID=ALIAS_ID). Then we create 5 ID fields (1 for each source plus a birth & current ID) for each record in the registry as shown below.

| Birth SV_ID | Birth Source | IRD Alias | IMM Alias | OTH Alias | Current ID |
|---|---|---|---|---|---|
| **SV_ID1** | IRD | SV_ID1 | | | SV_ID1 |
| **SV_ID2** | IRD | SV_ID2 | | | SV_ID2 |
| **SV_ID3** | IRD | SV_ID3 | | | SV_ID3 |
| •••• | •••• | •••• | •••• | •••• | •••• |
| **SV_ID4** | IMM | | SV_ID4 | | SV_ID4 |
| **SV_ID5** | IMM | | SV_ID5 | | SV_ID5 |
| **SV_ID6** | IMM | | SV_ID6 | | SV_ID6 |
| •••• | •••• | •••• | •••• | •••• | •••• |
| **SV_ID7** | OTH | | | SV_ID7 | SV_ID7 |
| **SV_ID8** | OTH | | | SV_ID8 | SV_ID8 |
| **SV_ID9** | OTH | | | SV_ID9 | SV_ID9 |
| •••• | •••• | •••• | •••• | •••• | •••• |

Current ID is defined as: Current_ID = Minimum (IRD_Alias, IMM_Alias, OTH_Alias) .

This will be a massive register that includes every entity record ever observed in the 3 sources. Now we start cross linking the 3 source entities and filling in the blank Alias IDs. Let us say that we discover that the IRD entity SV_ID3 and IMM entity SV_ID5 are the same entity. If this is the case we search the Birth_SV_ID for SV_ID3 and when we find that record we set IMM_ALIAS=SV_ID5. Then we search Birth_SV_ID for SV_ID5 and when we find that record we set IRD_ALIAS=SV_ID3. Our new structure becomes:

Administrative Data Paper    Version 4.6    35

| Birth SV_ID | Birth Source | IRD Alias | IMM Alias | OTH Alias | Current ID |
|---|---|---|---|---|---|
| **SV_ID1** | IRD | SV_ID1 | | | SV_ID1 |
| **SV_ID2** | IRD | SV_ID2 | | | SV_ID2 |
| **SV_ID3** | IRD | SV_ID3 | **SV_ID5** | | SV_ID3 |
| •••• | •••• | •••• | •••• | •••• | •••• |
| **SV_ID4** | IMM | | SV_ID4 | | SV_ID4 |
| **SV_ID5** | IMM | **SV_ID3** | SV_ID5 | | **SV_ID3** |
| **SV_ID6** | IMM | | SV_ID6 | | SV_ID6 |
| •••• | •••• | •••• | •••• | •••• | •••• |
| **SV_ID7** | OTH | | | SV_ID7 | SV_ID7 |
| **SV_ID8** | OTH | | | SV_ID8 | SV_ID8 |
| **SV_ID9** | OTH | | | SV_ID9 | SV_ID9 |
| •••• | •••• | •••• | •••• | •••• | •••• |

We remove duplicates with the rule Birth_SV_ID must equal Current_ID. This gives us a database that can be incrementally updated while preserving all previous information. Because our database preserves all the original birth records IDs, mistakes can be localized & fixed. In addition, the implied linkage database will contain none of the original source identifiers which may be confidential. Yet the internal administrative registers will allow us to obtain and use these identifiers if needed.

This large register mapping all inter-relations must exist but it need not be visible to most users. Instead, the "official person register" might be only those records where Birth_SV_ID=Current_ID. This is just possible way in which to create a true register with unique virtual identifiers.

## 10.3 MULTIPLE ENTITY TYPE REGISTERS

In addition to spanning multiple years, the BR encompasses multiple entity types. In general, the entity types must be closely related & hierarchical. In the BR, these might encompass two entity types: 1) enterprise entity and 2) location entity. Here it is assumed that the principle entity type is a legal business entity (enterprise) that operates in 1 or more locations. The figure below illustrates this fact. The key point here is the enterprise and locations must have separate unique sequential virtual identifiers (SV_ID) drawn from a single Virtual ID generator.

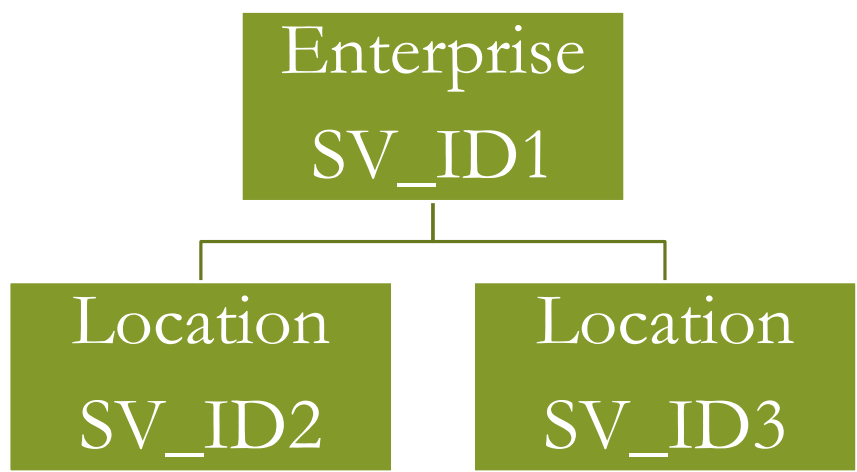

The different entity types can be perceived as coming from 2 fundamentally different source files but some of our definition may no longer be relevant. Following the previous logic we might create a file like this.

| Birth SV_ID | Birth Source | ENT Alias | LOC Alias |
|---|---|---|---|
| **SV_ID1** | ENT | SV_ID1 | |
| •••• | •••• | •••• | •••• |
| **SV_ID2** | LOC | SV_ID1 | SV_ID2 |
| **SV_ID3** | LOC | SV_ID1 | SV_ID3 |
| •••• | •••• | •••• | •••• |

How Statistics Canada does this is by creating the following 2 records. In the Statcan view one of the locations is always viewed as the head of a group of peers. Then there are complex rules for the definition of a unique location or a unique enterprise.

| Birth SV_ID | ENT Alias | LOC Alias |
|---|---|---|
| **SV_ID1** | SV_ID1 | SV_ID1 |
| **SV_ID2** | SV_ID1 | SV_ID2 |

## 10.4 SUMMARY

As one can see, each layer of complexity (source or entity type) increases the numbers of identifiers that must be appended to each record. Thus minimizing the number of entities covered by the register and the number of source files required to define the population is a good idea. In summary our virtual identifiers (SV_ID) must be unique and consistent across time, entity types and our core input sources for defining our various populations. In the context of the IDI, our register might span up to 3 types of entities: persons, family and household and 4 major sources (immigration, IRD, Births and other). All entity types must be collections of persons and the relationship must be hierarchical.

This document does not contain a detailed design for creating a unique sequential virtual identifier system. Rather, the document is intended to illustrate how one might create such a system.